# Four-hundred Very Metal-poor Stars Studied with LAMOST and Subaru. III. Dynamically Tagged Groups and Chemodynamical Properties

Ruizhi Zhang[1,2], Tadafumi Matsuno[3], Haining Li[1], Wako Aoki[4,5], Xiang-Xiang Xue[1], Takuma Suda[6,7], Gang Zhao[1,2], Yuqin Chen[1,2], Miho N. Ishigaki[4], Jianrong Shi[1], Qianfan Xing[1], and Jingkun Zhao[1]

[1] CAS Key Laboratory of Optical Astronomy, National Astronomical Observatories, Chinese Academy of Sciences, Beijing 100101, People's Republic of China; lhn@nao.cas.cn, gzhao@nao.cas.cn
[2] School of Astronomy and Space Science, University of Chinese Academy of Sciences, No.19(A) Yuquan Road, Shijingshan District, Beijing 100049, People's Republic of China
[3] Astronomisches Rechen-Institut, Zentrum für Astronomie der Universität Heidelberg, Mönchhofstraße 12-14, 69120 Heidelberg, Germany
[4] National Astronomical Observatory of Japan, 2-21-1 Osawa, Mitaka, Tokyo 181-8588, Japan; aoki.wako@nao.ac.jp
[5] Astronomical Science Program, The Graduate University for Advanced Studies, SOKENDAI, 2-21-1 Osawa, Mitaka, Tokyo 181-8588, Japan
[6] Department of Liberal Arts, Tokyo University of Technology, Nishi Kamata 5-23-22, Ota-ku, Tokyo 144-8535, Japan
[7] Research Center for the Early Universe, The University of Tokyo, 7-3-1 Hongo, Bunkyo-ku, Tokyo 113-0033, Japan

Received 2024 January 17; revised 2024 February 17; accepted 2024 February 21; published 2024 May 3

## Abstract

Very metal-poor (VMP) stars record the signatures of early accreted galaxies, making them essential tools for unraveling the early stages of Galaxy formation. Understanding the origin of VMP stars requires comprehensive studies of their chemical compositions and kinematics, which are currently lacking. Hence, we conduct a chemodynamical analysis of 352 VMP stars selected from one of the largest uniform high-resolution VMP star samples, jointly obtained from LAMOST and Subaru. We apply a friends-of-friends clustering algorithm to the master catalog of this high-resolution sample, which consists of 5778 VMP stars. It results in 131 dynamically tagged groups with 89 associated with known substructures in the Milky Way, including Gaia-Sausage-Enceladus (GSE), Thamnos, Helmi streams, Sequoia, Wukong, Pontus, and the very metal-poor disk (VMPD). Our findings are: (i) the VMPD shows lower Zn abundances than the rest, which indicates that it could be a relic of small stellar systems; (ii) Sequoia shows moderately high r-process abundances; (iii) Helmi streams show deficiencies in carbon and light neutron-capture elements; (iv) the fraction of carbon-enhanced metal-poor stars with no enhancement in heavy elements (CEMP-no stars) seems low in the VMPD and the Helmi streams; and (v) a subgroup in GSE exhibits a very high fraction of r-process enhanced stars, with four out of five showing [Eu/Fe]> +1.0. The abundance patterns of other elements in VMP substructures largely match the whole VMP sample. We also study large-scale correlations between abundance ratios and kinematics without classifying stars into substructures, but it does not yield significant correlations once the overall chemical evolution is considered for most elements.

*Unified Astronomy Thesaurus concepts:* Milky Way dynamics (1051); Milky Way evolution (1052); Milky Way Galaxy (1054); Galaxy chemical evolution (580); Galaxy abundances (574); Galactic archaeology (2178); Population II stars (1284)

*Supporting material:* machine-readable table

## 1. Introduction

Our Milky Way has gone through frequent mergers with dwarf galaxies during its formation, and these merging events have left abundant stellar substructures, mainly in the Galactic halo (Bullock et al. 2001; Zolotov et al. 2009; Purcell et al. 2010). Though relatively recent merging events can still be identified through spatial density in the form of streams or shells of stars (Ibata et al. 1994; Belokurov et al. 2006; Zhao et al. 2015; Ibata et al. 2021), debris left from more ancient accretions has already lost its spatial coherence. Fortunately, for relic stars that have originated from the same stellar system, the integrals of motion of their orbits within the Milky Way remain coherent for a much longer time (Helmi & White 1999; Helmi 2020), and hence we can excavate fossils of building blocks accreted a long time ago by identifying dynamical groups among Galactic halo stars. Moreover, elemental abundances in the atmospheres of stars can be used to trace the chemical evolution history of their progenitor stellar systems (Kobayashi et al. 2020). For example, the distribution of [α/Fe] with [Fe/H] reflects the star formation efficiency (Tinsley 1980) and has been widely used to distinguish stars that have been born in situ from those accreted from smaller systems (Nissen & Schuster 2010; Helmi et al. 2018; Matsuno et al. 2019, 2022a, 2022b; Fernandes et al. 2023; Horta et al. 2023).[8] Therefore, chemodynamical studies that combine dynamical properties and chemical abundances of halo stars enable us to identify various accretion components in the Galactic halo and uncover the nature of their progenitor systems.

Thanks to the combination of the Gaia mission (Gaia Collaboration et al. 2016, 2018, 2021) and various spectroscopic surveys including SEGUE from the Sloan Digital Sky Survey (Yanny et al. 2009), LAMOST (Zhao et al. 2006, 2012), GALAH (De Silva et al. 2015), APOGEE (Majewski et al. 2017), etc., the chemodynamical studies of the Milky Way halo stars have significantly advanced in recent years (e.g., Myeong et al. 2018a; Belokurov et al. 2018; Helmi et al. 2018; Koppelman et al. 2018; Helmi 2020; Naidu et al. 2020; Lövdal et al. 2022; Dodd et al. 2023). The combination of astrometry and radial-velocity measurements has enabled the identification of kinematic substructures, which are promising candidates for accreted galaxies. It is now clear that the Milky Way's inner stellar halo is dominated by debris of the last major merger Gaia-Sausage-Enceladus (GSE; Belokurov et al. 2018; Helmi et al. 2018) and those heated by this accretion from the disk present at that time (Splash; Belokurov et al. 2020). The nearby halo stars contain smaller but still quite significant kinematic substructures, such as the Helmi streams[9] (Helmi et al. 1999), Thamnos (Koppelman et al. 2019a), and Sequoia (Matsuno et al. 2019; Myeong et al. 2019), and there are also signatures of past accretion events in the outer part of the Galaxy as well, such as Cetus (Newberg et al. 2009) and Wukong/LMS-1 (Yuan et al. 2020a; Naidu et al. 2020). Although each kinematic substructure shows coherent chemical abundance trends among its member stars, the general trends can be different among different substructures (Aguado et al. 2021a; Matsuno et al. 2021, 2022a, 2022b; Horta et al. 2023). Thus, these differences enable detailed characterization of corresponding progenitor galaxies.

---

[8] The standard notations $[X/Y] = \log(N_X/N_Y) - \log(N_X/N_Y)_\odot$ and $A(X) = \log(N_X/N_H) + 12$ for elements X and Y are adopted in this work.

Very metal-poor (VMP) stars with [Fe/H]< −2.0 are low-mass old stars that preserve the record of the chemical composition and dynamics of the early Milky Way. Given their dominance in small dwarf galaxies (Simon 2019) and their potential association with disrupted stellar systems, VMP stars are expected to harbor a significantly higher fraction of accreted remnants from low-mass stellar systems compared to more metal-rich populations. Therefore, studying the chemodynamical properties of VMP stars is of great importance to fully reproduce the early history of our Galaxy, which involves the formation of in situ Galactic stars, accretions of low-mass dwarf galaxies such as the ultra-faint dwarf (UFD) galaxies, and those that have merged with the Milky Way in its very early stages.

However, despite recent progress, we still lack comprehensive chemodynamical studies on the VMP region of the Galactic halo. On the one hand, various chemodynamical studies on Galactic halo stars either focus on relatively higher metallicity regions or involve only a very limited number of VMP stars (e.g., Monty et al. 2020; Aguado et al. 2021a, 2021b; Limberg et al. 2021b; Gull et al. 2021; Matsuno et al. 2022a, 2022b; Carrillo et al. 2022; da Silva & Smiljanic 2023; Horta et al. 2023; Limberg et al. 2024). On the other hand, a few studies have examined the dynamical properties of VMP stars but have not provided detailed information on elemental abundances (e.g., Yuan et al. 2020b, hereafter Y20; Limberg et al. 2021a; Carollo et al. 2023).

Chemodynamical analyses of VMP stars hold the key to unraveling early nucleosynthesis and the origins of diverse elements, thanks to their minimal enrichment from subsequent generations of stars. Low-metallicity stars in the halo show diverse chemical abundance ratios, especially in neutron-capture elements (e.g., McWilliam et al. 1995; Honda et al. 2004; Barklem et al. 2005). Such a dispersion can be due to inhomogeneous mixing of the ejecta from nucleosynthesis events (e.g., Argast et al. 2004; Hirai et al. 2015) and/or differences in the chemical enrichments among different galaxies (Ishimaru et al. 2015; Ojima et al. 2018). Therefore, separating stars according to their progenitor galaxy whenever possible is highly beneficial in observationally disentangling the two mechanisms. Moreover, if one can associate an abundance difference between two galaxies to their different properties, such as mass or star formation timescale, it allows us to constrain the property of the nucleosynthesis process producing the elements.

We surpass previous chemodynamical analyses of VMP stars by investigating a large, homogeneous sample selected from the LAMOST survey, leveraging the high-resolution spectroscopy using the High Dispersion Spectrograph (HDS; Noguchi et al. 2002) equipped on the Subaru telescope. This allows us to identify distinct groups of stars sharing similar dynamics within the Galactic halo with a technique known as dynamical tagging. This paper is the third one of a series. Detailed descriptions of target selection and observations of this LAMOST/Subaru VMP sample are provided in (Aoki et al. (2022, hereafter Paper I), which also reports radial velocities and interstellar reddening of the program stars. A homogeneous chemical abundance analysis for this LAMOST/Subaru VMP star sample is presented in Li et al. (2022, hereafter Paper II), where stellar parameters and abundances for more than 20 species have been determined for 385 stars in the LAMOST/Subaru VMP sample.

This paper is organized as follows. In Section 2, we summarize the properties of the sample and the determination of their basic kinematic parameters. The method of dynamical analysis and the corresponding result of dynamical clustering of the sample are respectively presented in Sections 3.1 and 3.2. We discuss chemodynamical properties of identified VMP substructures in Section 4.1, kinematics of stars with peculiar abundance ratios in Section 4.2, and large-scale chemodynamical correlations in Section 4.3. We present the conclusion in Section 5.

## 2. Data

Although the VMP sample provided by Paper II allows us to investigate the chemodynamics of the Milky Way halo in a very low metallicity region, its size is not sufficiently large compared to those used to define kinematic substructures or dynamically tagged groups (DTGs). Hence, we make use of the LAMOST 10,000 VMP star catalog by Li et al. (2018, hereafter L18) to construct a larger VMP sample, which can be regarded as the master catalog for the high-resolution VMP sample. This large sample allows us to more accurately define DTGs, which we can then use to associate VMP stars from Paper II to corresponding DTGs. In the following discussion, we refer to the high-resolution VMP sample from the LAMOST/Subaru survey as the "HR sample" and the larger VMP master sample from LAMOST as the "LR sample."

### 2.1. Kinematics of the HR Sample

Candidates for the HR sample were originally selected from LAMOST DR3 through DR5 and were then follow-up observed using the HDS on the Subaru telescope. We have measured 1D local thermodynamic equilibrium (LTE) abundances of Li, C, Na, Mg, Si, Ca, Sc, Ti, V, Cr, Mn, Co, Ni, Zn, Sr, Y, Zr, Ba, La, and Eu based on the high-resolution spectra and applied a non-LTE correction to Na abundances. Details about the target selection, follow-up strategy, and abundance analysis can be referred to in Papers I and II. An advantage of our sample is that the spectra have been analyzed homogeneously. Therefore, our chemodynamical investigation presented in the present paper is not subject to significant systematic uncertainties.

---

[9] Because of the bimodal component in vertical velocity, we refer to this structure as streams (see Helmi 2020 and references therein).

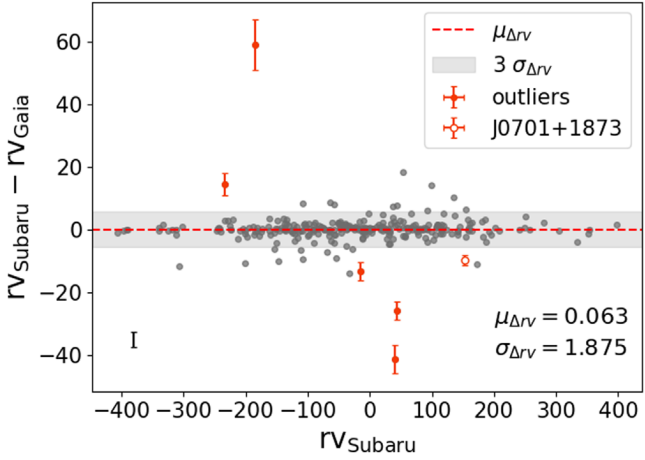

**Figure 1.** The distributions of the differences between Gaia and Subaru radial-velocity measurements. The red dashed line and the gray shaded region are, respectively, the median value and $3\sigma$ scale of the residuals. The red open dot is J0701 + 1813, which has consistent radial velocities between Subaru and LAMOST DR9, and the red solid dots are sources with inconsistent radial velocities between Subaru spectra and Gaia DR3. The median and dispersion of the differences are shown in the lower right of this panel, and the typical error bar of the whole sample is shown in the lower left.

We cross-matched this HR sample with the Gaia DR3 catalogs (Gaia Collaboration et al. 2023) and the distance catalog of Bailer-Jones et al. (2021, hereafter BJ21). All of our 385 stars have parallax, proper-motion, and Gaia magnitude measurements, and photogeometric distances from BJ21. In this paper, we primarily use the photogeometric distances when computing stellar kinematics. Of these photogeometric distances, only ten stars have a relative distance uncertainty larger than 20%, which is defined as $(d_{\rm hi,photogeo} - d_{\rm lo,photogeo})/(2 \times d_{\rm med,photogeo})$, where $d_{\rm hi,photogeo}$, $d_{\rm lo,photogeo}$, and $d_{\rm med,photogeo}$ are the 84th and 16th percentiles, and median of BJ21 photogeometric distance, respectively. We excluded these ten stars from the kinematic analysis, and the median relative distance uncertainty of the rest of the HR sample is about 3%. We further removed 18 stars with the renormalized unit weighted error (RUWE) larger than 1.4 since their astrometry is less reliable (Lindegren et al. 2021).

In addition to astrometry, we also needed radial-velocity measurements to compute kinematics. In Figure 1, we compare the radial velocities reported in Paper I based on Subaru spectra[10] with the radial velocities provided by Gaia DR3 for 273 stars whose Gaia radial velocities are available. While the median uncertainty in the Gaia radial velocity ($\sigma_{\rm Gaia}$) is 2.1 km s$^{-1}$, we adopt 1.0 km s$^{-1}$ as the typical uncertainty in the Subaru radial velocity ($\sigma_{\rm Subaru}$), considering the long-term stability of HDS. The median velocity difference ($\mu_{\Delta rv}$) and half the difference between the 16th and 84th percentiles ($\sigma_{\Delta rv}$) are 0.063 km s$^{-1}$ and 1.875 km s$^{-1}$, respectively. Given the velocity uncertainties, we conclude that these median offset and dispersion values are small, indicating that the two velocities are consistent.

However, some outliers need further discussion. We treated six stars with $|{\rm rv}_{\rm Subaru} - {\rm rv}_{\rm Gaia} - \mu_{\Delta rv}| > 3\sqrt{\sigma_{\rm Subaru}^2 + \sigma_{\rm Gaia}^2 + \sigma_{\Delta rv}^2}$ as outliers. We cross-matched these stars with LAMOST DR9 to check whether they have variable radial velocities. One star (J0701 +1813) has similar radial velocities between Subaru and LAMOST measurements at different observation dates; thus, we consider the Subaru measurement reliable. The other five stars have different radial velocities among Subaru spectra, Gaia, and LAMOST, and three of the stars show clear radial-velocity variation among the multiple observations from LAMOST. We removed these five stars since they are likely part of binary systems and we do not have reliable radial velocities of the system's barycenters. In total, the final HR sample selected for chemodynamical analysis consists of 352 stars with reliable astrometry and radial-velocity measurements.

We computed the positions and velocities of the stars adopting the photogeometric distance, coordinates, and proper motion from the Gaia DR3, and the radial velocity derived from Subaru spectra. We assumed that the Sun is located at 8.21 kpc from the Galactic Center (McMillan 2017) and 20.8 pc above the Galactic plane (Bennett & Bovy 2019). We took the local standard of rest (LSR) velocity as $V_{\rm LSR} = 233.1$ km s$^{-1}$ from the measurement by McMillan (2017) and adopted the solar peculiar motion $(U_\odot, V_\odot, W_\odot) = (11.1, 12.24, 7.25)$ km s$^{-1}$ as measured by Schönrich et al. (2010). We used a right-hand Galactocentric Cartesian coordinate system $(X, Y, Z)$ and a Galactocentric cylindrical coordinate system $(R, \phi, z)$, where the Sun is at $(X_\odot, Y_\odot, Z_\odot) = (-8.21, 0.0, 0.0208)$ kpc, and $v_\phi$ is negative for stars on the disk rotation.

Using the software Agama (Vasiliev 2019), we further computed the following quantities that characterize the orbits of the stars and integrated five times the period of the circular orbit corresponding to the orbital energy for each star to derive the orbital parameters:

1. $E = 1/2(v_R^2 + v_\phi^2 + v_z^2) + \Psi_{\rm MW}(R, z)$,
2. $L_\perp = \sqrt{(Rv_z - zv_R)^2 + (zv_\phi)^2}$,
3. $J_\phi = -L_z = -Rv_\phi$, where a positive azimuthal action indicates a prograde orbit, which is opposite for the z-component of angular momentum,
4. radial action ($J_r$) and vertical action ($J_z$),
5. total action $J_{\rm tot} = J_r + J_z + |J_\phi|$,
6. circularity $\eta = J_\phi/|L_{z,{\rm max}}(E)|$, where $L_{z,{\rm max}}$ is the angular momentum of the circular orbit with orbital energy $E$,
7. $R_{\rm max}$ and $|z|_{\rm max}$,
8. Galactocentric distance at pericenter ($r_{\rm peri}$) and apocenter ($r_{\rm apo}$),
9. eccentricity ($ec = \frac{r_{\rm apo} - r_{\rm peri}}{r_{\rm apo} + r_{\rm peri}}$) and semimajor axis ($a = \frac{r_{\rm apo} + r_{\rm peri}}{2}$) of the orbit.

We assumed the axisymmetric Milky Way potential of McMillan (2017) for $\Psi_{\rm MW}(R, z)$. For each star, we ran a 10,000 times Monte Carlo (MC) simulation assuming Gaussian uncertainties for the observables to estimate uncertainties for these quantities. We note that, although the distribution of the BJ21 photogeometric distance does not follow a Gaussian distribution, over 90% of the stars in the HR sample show almost symmetric confidence intervals with $|(\sigma_{\rm hi} - \sigma_{\rm lo})/(\sigma_{\rm hi} + \sigma_{\rm lo})| < 20\%$, where $\sigma_{\rm hi} = d_{\rm hi,photogeo} - d_{\rm med,photogeo}$ and $\sigma_{\rm lo} = d_{\rm med,photogeo} - d_{\rm lo,photogeo}$.

We note here that the interpretation of the observed correlations between kinematics and chemical abundances might need caution because of our sample selection (see Section 2.2 of Paper I). While we focused on bright stars at [Fe/H] > −3, we extended the sample to fainter stars at [Fe/H] < −3. This observing strategy introduces a correlation between metallicity and heliocentric distance (Figure 2), which can then introduce a correlation between metallicity and kinematic quantities. Since abundance ratios evolve with metallicity, the metallicity–kinematics correlation would further lead to an apparent correlation between chemical abundance ratios and kinematics. We try to mitigate this effect throughout the paper.

---

[10] We note that the radial velocity for a few objects from Subaru spectra reported in Paper I needs further corrections. We use the corrected values in this work, for which readers can refer to the erratum of Paper I.

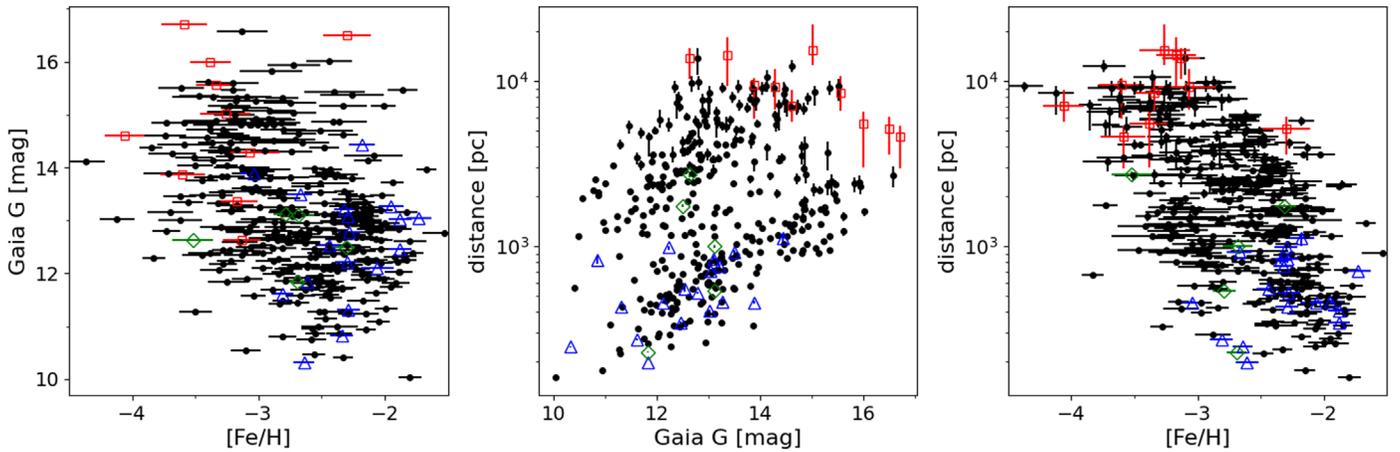

**Figure 2.** The relations between Gaia $G$-band magnitude, photogeometric distance (Bailer-Jones et al. 2021), and metallicity from Paper II of the stars. Red open squares, blue open triangles, and green open diamonds, respectively, show stars having a relative distance uncertainty larger than 20%, stars having RUWE larger than 1.4, and those having an inconsistent radial velocity between Subaru spectra, Gaia DR3, and LAMOST DR9. These figures show a clear correlation between distance and [Fe/H] because of our high-resolution observation strategy described in Paper I.

### 2.2. Kinematics of the LR Sample

We combined the L18 catalog with our HR sample to construct a larger VMP master sample for clustering. To avoid the influence of the relatively metal-rich stars in clustering, we cross-matched the L18 catalog with LAMOST DR9 and removed stars with signal-to-noise ratios (S/N) less than 15 or [Fe/H] > −1.5, to eliminate the stars that are incorrectly classified as VMP in L18. Note that the official LAMOST pipeline does not provide metallicity estimation at [Fe/H] < −2.5, and thus we did not exclude stars with no parameter estimation in LAMOST DR9, where no metallicity estimation may also imply a relatively low metallicity.

Subsequently, we cross-matched the L18 sample with Gaia DR3 and the distance catalog of BJ21, and obtained a sample of 7741 VMP stars with Gaia astrometric data. We excluded 239 stars with RUWE > 1.4. Additionally, we removed two stars located in a crowded field, where there is more than one star within a 5″ radius, which may cause incorrect cross-matching. Using the Gaia source id, we also found 249 duplicate sources, and we only kept the data with the highest LAMOST S/N for each of them and eliminated 260 records. In addition, 1020 stars with a reliable radial velocity from neither LAMOST DR9 nor Gaia DR3 were also removed.

There are ∼500 faint and distant giants in the reduced sample. For some of these giants, the photogeometric distances from BJ21 may be underestimated (see, e.g., Section 5.5 in BJ21), and the parallaxes may not be accurate enough for distance determinations. Therefore, we introduced distances from Zhang et al. (2023, hereafter Z23) for such giants, who derived the distances of ∼20,000 K giants in the Galactic halo from photometry following a procedure similar to that of Xue et al. (2014). Since distances from Z23 and BJ21 might not be on the same scale, we applied a correction to Z23. We first selected giants with $\log g < 3.5$ and cross-matched them with the K-giant catalog from Z23, and then fit a linear relation between Z23 and BJ21 distances, assuming that stars follow a normal distribution around the relation. We applied a Markov Chain Monte Carlo (MCMC) method to fit the relation to the 111 VMP giants within 4 kpc ($\varpi > 0.25$ mas). Figure 3 shows the best-fitting result and we calibrated Z23 distances using this fit relation.

A criterion is needed for different distance adoptions. Since giants with lower surface gravity are usually more luminous, stars with lower $\log g$ would have smaller apparent magnitudes at the same distance. We present all the stars with calibrated Z23 distances greater than 5 kpc on the $\log g - G_{\mathrm{Gaia}}$ panel as shown in Figure 4, where $\log g$ are from LAMOST DR9, and $G_{\mathrm{Gaia}}$ is the $G$-band apparent magnitude from Gaia DR3. We found that these distant giants tend to be above the black line in Figure 4 and took this black line as the division of different distance estimates. We used calibrated Z23 distances for stars within the upper-right area, while, out of this area, the BJ21 photogeometric distances are adopted.

We then removed 545 stars with large distance uncertainties ($d_{\mathrm{err}}/d > 20\%$). Among the remaining 5675 stars, we adopted BJ21 distances for 5444 objects and Z23 distances for 231 objects. The typical uncertainties are 6% and 14% for distances from BJ21 and Z23, respectively. This reduced large VMP sample has a similar distance distribution to the HR sample as shown in Figure A1.

We used the same coordinate systems and computed the same parameters for the selected L18 catalog as for the HR sample. For radial velocity, we prioritized the measurement from LAMOST DR9; otherwise, we adopted the value from Gaia DR3. We found two unbound stars with positive energy and removed them from the following analysis. Unlike other works that aim to explore the clustering substructures of VMP stars in the Galactic halo (e.g., Y20; Limberg et al. 2021a), we did not make any additional restrictions in the Toomre diagram, which allows us to investigate the VMP disk-like substructures.

For 247 common stars between the HR sample and the L18 catalog, we replaced the kinematic parameters of these stars in the L18 catalog with those computed in Section 2.1 and combined these two tables. Finally, we obtained the LR sample of 5778 VMP stars with kinematic parameters for clustering.

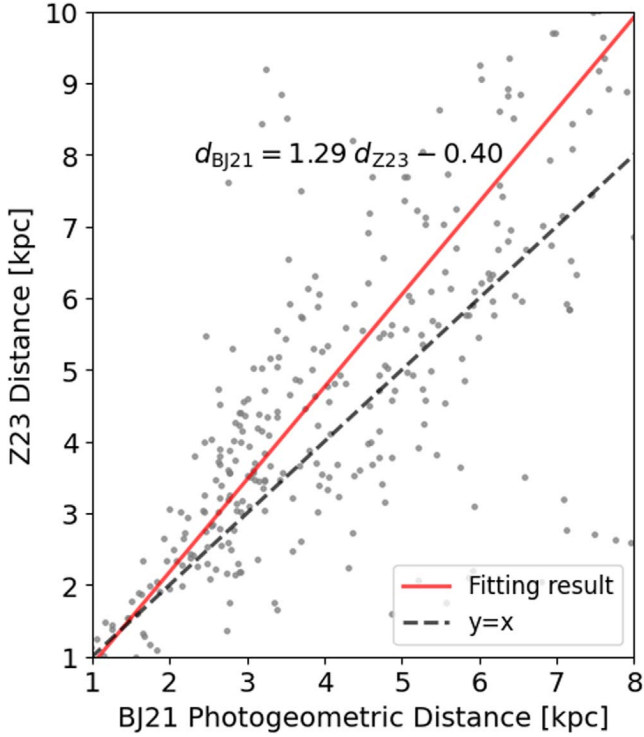

**Figure 3.** The comparison of BJ21 and Z23 distances for the VMP giants with $d_{Z23,\text{error,calib}}/d_{Z23,\text{calib}} < 0.2$ and $(d_{\text{hi,photogeo}} - d_{\text{lo,photogeo}})/(2 \times d_{\text{med,photogeo}}) < 0.2$. The red line is the fitting result using MCMC.

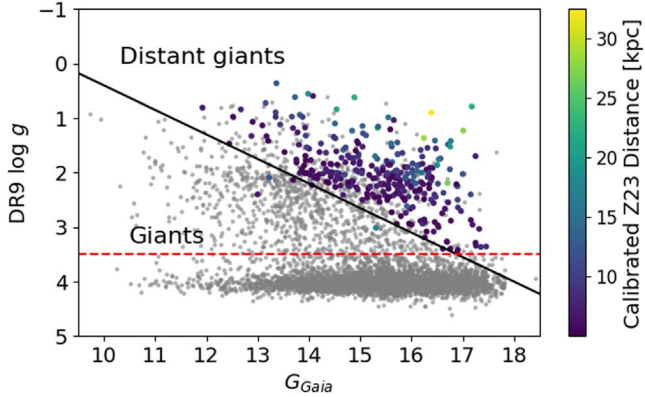

**Figure 4.** The distribution of $G$-band apparent magnitude and surface gravity of stars in the L18 catalog. The colored dots are stars with calibrated Z23 distances greater than 5 kpc. The red dashed line is $\log g = 3.5$, and the black line is the criterion we adopted to select distant giants. The upper-right region beyond these two lines is the area where calibrated Z23 distances are used for stars.

To summarize, we constructed two samples with reliable dynamical parameters based on the LAMOST/Subaru high-resolution survey and the low-resolution sample from LAMOST, respectively. We emphasize that we only used the kinematics of the LR sample for clustering and the chemodynamical analysis will focus on the HR sample.

## 3. Clustering Analysis

### 3.1. Method

Integrals of motion (IoM) of stars originating from the same progenitor system remain coherent over time in a slowly varying gravitational potential, implying that stars with similar IoMs may share a common origin. An axisymmetric system usually admits three isolating IoMs, and under the action-angle coordinates the three actions ($J_r$, $J_\phi$, $J_z$) are useful isolating integrals to characterize the stellar orbits in axisymmetric galaxies in dynamical equilibrium (Ollongren 1965; Binney & Tremaine 2008; Binney 2012). Helmi & White (1999) first proposed that disrupted satellites could be identified as clusters in the action-angle phase space. In the Gaia era, many substructures and DTGs are found both in action phase space and in energy–action combined phase space (Myeong et al. 2018b; Roederer et al. 2018; Myeong et al. 2019; Limberg et al. 2021a; Shank et al. 2022a, 2022b; Lövdal et al. 2022; Cabrera Garcia et al. 2024; Dodd et al. 2023), which demonstrates the feasibility of searching for substructures in action phase space.

In this work, we used the friends-of-friends (FoF) clustering algorithm to identify substructures in the ($J_r$, $J_\phi$, $J_z$) phase space. This algorithm links two stars with a separation $\delta$ smaller than a given threshold (linking length) into one group, and this group continues to grow until the $\delta$ between any star in this group and all the other nongroup stars is greater than the linking length.

We adopted the following definition of $\delta$, which is similar to that of Wang et al. (2022) but in a different phase space.

$$\delta_{ij}^2 = \omega_{\Delta\sqrt{J_r}}(\sqrt{J_{r,i}} - \sqrt{J_{r,j}})^2 + \omega_{\Delta J_\phi}(J_{\phi,i} - J_{\phi,j})^2 + \omega_{\Delta\sqrt{J_z}}(\sqrt{J_{z,i}} - \sqrt{J_{z,j}})^2. \quad (1)$$

For radial and vertical actions, we considered separations in their square roots. This helped us identify clusters on polar or eccentric orbits, as $J_z$ and $J_r$ of stars on such orbits can vary significantly when their orbits are changed slightly.

The $\omega_{\Delta\sqrt{J_r}}$, $\omega_{\Delta J_\phi}$, $\omega_{\Delta\sqrt{J_z}}$ are weights used to normalize the separations in the three different actions. We normalized each action so that the median deviations of all the pairs are 1, i.e.,

$$\omega_{\Delta X} = \frac{1}{\text{median}[(\Delta X)^2]} \quad (X = \sqrt{J_r}, J_\phi, \sqrt{J_z}). \quad (2)$$

In our sample, the $\omega_{\Delta X}$ are $1/8.34 \text{ s}^{1/2} \text{ kpc}^{-1/2} \text{ km}^{-1/2}$, $1/821.74 \text{ s kpc}^{-1} \text{ km}^{-1}$, and $1/6.98 \text{ s}^{1/2} \text{ kpc}^{-1/2} \text{ km}^{-1/2}$ for $\sqrt{J_r}$, $J_\phi$, and $\sqrt{J_z}$, respectively.

To ensure that we selected reliable groups, for each group identified by the FoF algorithm we chose a suitable linking length, as shown in Appendix B. Only groups with no less than five members were kept for dynamic analysis.

### 3.2. Results

Based on the clustering results of the LR sample, we identified 286 DTGs; for 131 of them, members can be found in the HR sample. The result of the clustering analysis is shown in Table 1, where we also show the median dynamical parameters for each group. We found that 89 out of the 131 DTGs can be associated to known substructures in the Milky Way halo, namely, the GSE (Belokurov et al. 2018; Helmi et al. 2018), Thamnos (Koppelman et al. 2019a), Sequoia/I'itoi (Myeong et al. 2019; Naidu et al. 2020), Helmi streams (Helmi et al. 1999), Pontus (Malhan 2022), and Wukong/LMS-1 (Yuan et al. 2020a; Naidu et al. 2020). Table 2 lists such associations of our DTGs with the substructures, mostly following the criteria shown in Table C1.[11] We note that we relaxed the criteria for these associations to have a sufficient

---
[11] For groups that are close to substructures but do not meet the criteria, we decided whether they should be assigned to these substructures manually.

**Table 1**
Dynamical Parameters of DTGs Clustered by FoF Algorithm

| Group | Linking Length | $N_{LR}$[a] | $N_{HR}$[b] | $E$<br>($\sigma_E$)<br>($10^5$ km$^2$ s$^{-2}$) | $(J_r, J_\phi, J_z)$<br>$(\sigma_{J_r}, \sigma_{J_\phi}, \sigma_{J_z})$<br>(kpc km s$^{-1}$) | $(v_R, v_\phi, v_z)$<br>$(\sigma_{v_R}, \sigma_{v_\phi}, \sigma_{v_z})$<br>(km s$^{-1}$) | $ec$<br>$\sigma_{ec}$ | $\|z\|_{max}$<br>$\sigma_{\|z\|_{max}}$<br>(kpc) |
|---|---|---|---|---|---|---|---|---|
| DTG-1 | 0.098 | 212 | 6 | −1.738<br>0.045 | (501.9, 427.3, 96.0)<br>(93.4, 125.3, 25.9) | (−3.59, −49.74, −1.93)<br>(87.79, 17.62, 66.91) | 0.789<br>0.073 | 3.518<br>0.644 |
| DTG-2 | 0.112 | 105 | 3 | −1.659<br>0.041 | (251.3, 1087.4, 61.8)<br>(68.9, 158.7, 12.1) | (31.60, −127.37, −11.01)<br>(83.11, 21.07, 53.26) | 0.494<br>0.067 | 2.047<br>0.268 |
| DTG-3 | 0.098 | 93 | 3 | −1.733<br>0.043 | (704.8, -23.0, 74.6)<br>(52.6, 126.2, 28.1) | (−27.98, 2.70, −6.38)<br>(85.16, 14.40, 51.95) | 0.893<br>0.048 | 3.913<br>0.431 |
| DTG-4 | 0.100 | 82 | 5 | −1.736<br>0.062 | (603.2, 377.3, 28.7)<br>(62.6, 124.8, 9.1) | (−36.10, −44.63, −3.44)<br>(101.08, 13.38, 36.08) | 0.817<br>0.037 | 1.352<br>1.371 |
| DTG-5 | 0.134 | 79 | 3 | −1.702<br>0.060 | (589.5, 25.9, 301.4)<br>(82.5, 135.4, 50.2) | (−40.2, −4.46, 57.15)<br>(86.76, 17.17, 104.63) | 0.891<br>0.042 | 7.175<br>1.150 |

**Notes.** The full table in machine-readable form is available. The value and uncertainty of each parameter is the median value and half the difference between the 16th and 84th quantiles of each group.
[a] $N_{LR}$ is the number of stars from the LR sample in the DTG.
[b] $N_{HR}$ is the number of stars from the HR sample in the DTG.

(This table is available in its entirety in machine-readable form.)

number of HR stars for each group for later investigations on chemical properties.

We also found some DTGs with thick-disk-like kinematics with $|z|_{max} < 3$ kpc and $0.25 < ec < 0.60$. Although they occupy a similar region as the thick-disk stars in the Toomre diagram, their rotational velocities peak at roughly $-130$ km s$^{-1}$, slower than the typical rotational velocity of the canonical thick disk ($v_\phi \sim -180$ km s$^{-1}$; Carollo et al. 2010). Thus, we call these DTGs the very metal-poor disk (VMPD) throughout this paper.

Figure 5 shows the energy–action distribution of the identified substructures. We present the distributions of the LR sample and HR sample in the upper and lower panels of Figure 5, respectively. In addition to these substructures, we also found smaller DTGs, which we could not associate with known substructures. Among these new DTGs, we discuss the four groups containing more than ten members and at least three stars in the HR sample. The dynamical properties of these four groups are also shown in Figure 6 and Table 2. The dynamical parameters and clustering results of the HR sample are available in Zenodo and can be accessed via doi:10.5281/zenodo.10780897.

Again, we stress that we allowed some contamination by adopting relatively loose criteria so that we have sufficient numbers of stars for chemical investigation with the HR sample. Although we did not consider that several outliers would alter the interpretation of chemical abundance, caution is needed when the interpretation is based on a small number of stars.

## 4. Chemodynamics of VMP Stars

### 4.1. Early Chemical Evolution of Substructures and DTGs

In this section, we combine the chemical abundances of 352 stars in the HR sample as introduced in Section 2.1 with the clustering results from Section 3.2 to present the chemical patterns of carbon, $\alpha$-elements (Mg, Si, Ca, Ti), light odd-$Z$ elements (Na, Sc), iron-peak elements (V, Cr, Mn, Co, Ni, Zn), and heavy elements (Sr, Y, Zr, Ba, La, Eu) for each substructure.

The purpose of our study is to investigate the chemical evolution of the progenitors of the substructures, assuming that VMP stars retain the chemical properties of the interstellar medium from which they have formed. Any stars that have experienced external contamination to their surface chemical abundance should be excluded from our analysis. For example, the chemical compositions of carbon-enhanced metal-poor (CEMP) stars with large excesses of s-process elements (CEMP-s) are changed due to mass transfer from their companions (Abate et al. 2015; Jorissen et al. 2016), and thus we excluded ten CEMP-s stars and a potential CEMP-s star J0446 + 2124 (see Section 4.2.1 for details) when discussing the abundance distribution.

To make sure that our studies are based on the most reliable chemical abundances, we excluded stars with uncertainties larger than 0.2 dex in [Fe/H] and [X/Fe] during discussions on corresponding element X. Two stars with metallicity larger than −1.5 were also eliminated as their metallicities are much higher than the rest of the sample.

Using the selected HR sample, we present the results for GSE, VMPD, retrograde substructures, polar substructures, and other DTGs, respectively.

We also utilize the elemental abundances from the literature and the GALAH survey to compare the chemical distribution between our results and previous studies in a wider metallicity range. Readers may refer to Appendix D for details.

#### 4.1.1. GSE

GSE is the remnant of a major merger event of the ancient Milky Way, which was revealed by the "sausage"-like distribution in $v_R - v_\phi$, extremely radial/eccentric orbits, and low $\alpha$-element abundances of its members (Belokurov et al. 2018; Haywood et al. 2018; Helmi et al. 2018). Its progenitor merged with the Milky Way approximately 10 Gyr ago (Helmi et al. 2018; Gallart et al. 2019; Helmi 2020; Montalbán et al. 2021) and dominates the inner Galactic halo (e.g., see Lancaster et al. 2019; Naidu et al. 2020; Bird et al. 2021; Wu et al. 2022).

GSE is the largest substructure we recovered, taking up one-third of the LR sample and one-fourth of the HR sample. This result is similar to previous studies on the VMP halo using clustering algorithms (Limberg et al. 2021a; Shank et al. 2022b, 2022a) and Gaussian mixture models (An & Beers 2021).

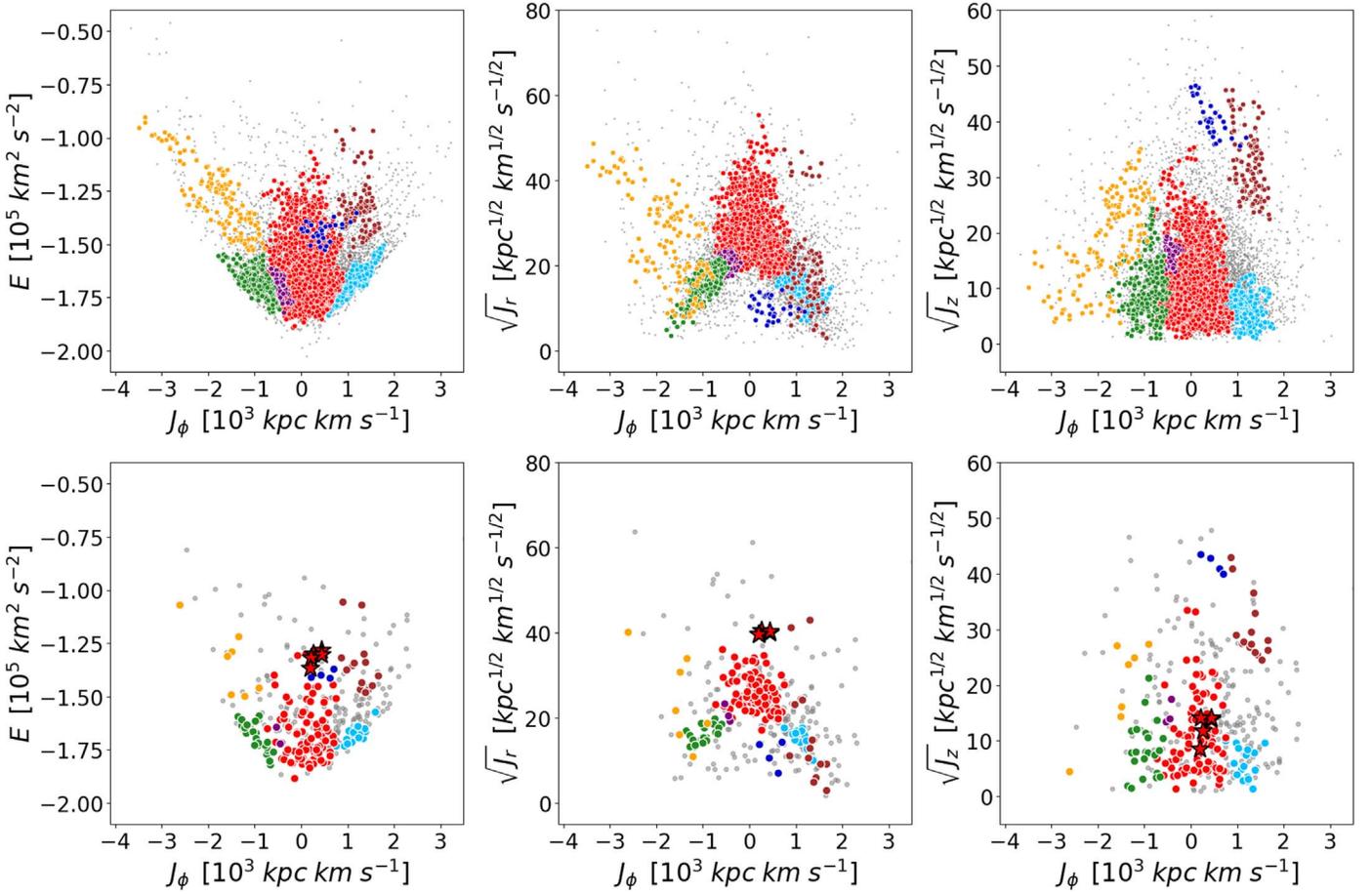

**Figure 5.** The distributions of groups associated with known substructures in energy and action phase space. The upper and lower panels are the results for the LR sample and the HR sample, respectively. Different colors represent different substructures: GSE (red), Thamnos (green), Sequoia (orange), Helmi streams (brown), Pontus (purple), Wukong/LMS-1 (blue), and VMPD (sky blue). The red stars stand for the $r$-process-enhanced subgroup in GSE; see Section 4.2.3 for details.

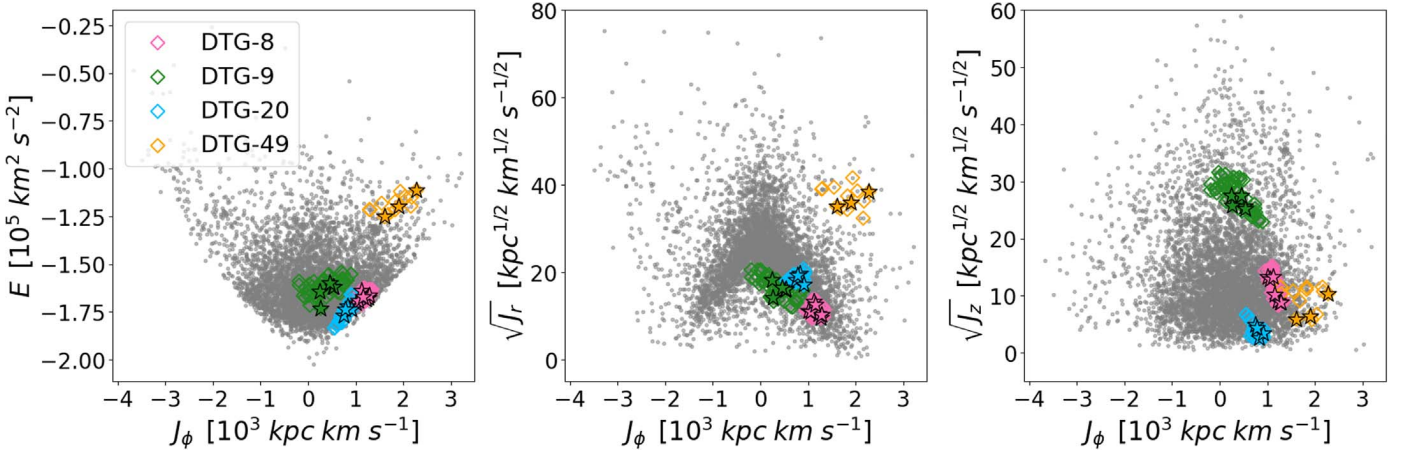

**Figure 6.** The distributions of four large DTGs in energy and action phase space. The open diamond symbols are group members in the LR sample, and the star symbols with black borders are stars in the HR sample. Different colors represent different DTGs: DTG-8 (pink), DTG-9 (green), DTG-20 (blue) and DTG-49 (orange).

The chemical distribution of GSE is shown in Figure 7. There is no significant difference between the distribution of GSE and that of the entire sample. This may be due to the fact that the chemical evolution of the massive progenitor of GSE ($M_\star \sim 10^8$–$10^9 M_\odot$; see Mackereth et al. 2019; Vincenzo et al. 2019; Helmi 2020; Naidu et al. 2020, 2021 and references therein) and the Milky Way are similar at [Fe/H] $< -2$. Another noticeable feature of GSE is the large scatter in most elements (e.g., C, Mg, Ca, Cr, and heavy elements). This property suggests that the GSE possesses a complicated chemical composition, which is also reported in several previous works (e.g., Zhao & Chen 2021; Donlon & Newberg 2023). This may indicate a series of merging processes between the massive progenitor of GSE and other dwarf galaxies before it merged with our Galaxy.

Table 2
Dynamical Parameters of Known Substructures and Large New DTGs

| Substructure | Linking Length | $N_{LR}$ | $N_{HR}$ | $E$ $(\sigma_E)$ $(10^5 \text{ km}^2 \text{ s}^{-2})$ | $(J_r, J_\phi, J_z)$ $(\sigma_{J_r}, \sigma_{J_\phi}, \sigma_{J_z})$ $(\text{kpc km s}^{-1})$ | $(v_R, v_\phi, v_z)$ $(\sigma_{v_R}, \sigma_{v_\phi}, \sigma_{v_z})$ $(\text{km s}^{-1})$ | $ec$ $\sigma_{ec}$ | $|z|_{max}$ $\sigma_{|z|_{max}}$ (kpc) |
|---|---|---|---|---|---|---|---|---|
| GSE | [0.056, 0.350] | 1940 | 89 | −1.684 0.131 | (649.9, 113.8, 89.1) (263.4, 376.9, 120.1) | (−18.25, −13.84, −1.34) (146.03, 43.56, 69.37) | 0.857 0.093 | 4.202 3.013 |
| VMPD | [0.072, 0.214] | 336 | 18 | −1.671 0.047 | (235.6, 1087.8, 55.0) (95.2, 253.4, 32.4) | (19.98, −129.49, −3.33) (78.43, 30.78, 46.17) | 0.490 0.117 | 1.876 0.787 |
| Thamnos | [0.118, 0.306] | 282 | 23 | −1.693 0.060 | (266.6, −881.1, 75.9) (107.2, 270.5, 82.3) | (−3.15, 107.24, −1.06) (74.75, 32.05, 58.71) | 0.544 0.126 | 2.462 1.445 |
| Sequoia | [0.202, 0.682] | 136 | 7 | −1.283 0.169 | (727.8, −1687.0, 388.5) (640.7, 666.5, 330.2) | (40.94, 199.52, −32.96) (187.35, 80.51, 146.77) | 0.590 0.163 | 9.484 4.465 |
| Helmi streams | [0.290, 0.628] | 112 | 15 | −1.329 0.081 | (258.7, 1331.4, 991.4) (230.8, 260.3, 381.2) | (10.97, −153.11, −186.19) (114.98, 34.97, 234.00) | 0.379 0.152 | 14.308 4.060 |
| Pontus | [0.124, 0.158] | 62 | 3 | −1.702 0.044 | (429.8, −435.1, 251.6) (77.7, 88.0, 62.9) | (−47.1, 54.03, −71.25) (85.88, 10.79, 94.65) | 0.734 0.053 | 5.401 0.825 |
| Wukong (LMS-1) | [0.308, 0.366] | 30 | 4 | −1.431 0.043 | (98.5, 432.0, 1597.3) (58.1, 275.0, 322.3) | (29.7, −43.60, 171.83) (103.24, 46.79, 202.93) | 0.263 0.065 | 12.397 1.265 |
| DTG-8 | 0.114 | 60 | 7 | −1.665 0.024 | (130.3, 1150.4, 132.8) (29.1, 90.9, 47.8) | (24.38, −140.66, −8.98) (62.94, 11.43, 77.68) | 0.368 0.038 | 3.124 0.775 |
| DTG-9 | 0.212 | 53 | 5 | −1.611 0.040 | (270.2, 343.1, 790.5) (80.3, 340.8, 150.8) | (9.22, −40.51, −126.87) (85.23, 39.03, 154.08) | 0.541 0.091 | 9.284 0.972 |
| DTG-20 | 0.126 | 32 | 4 | −1.756 0.042 | (348.8, 756.8, 14.3) (29.8, 124.3, 9.8) | (15.45, −89.52, 6.02) (53.80, 9.83, 28.79) | 0.627 0.022 | 0.803 0.275 |
| DTG-49 | 0.476 | 13 | 3 | −1.197 0.037 | (1400.2, 1829.9, 107.2) (171.4, 317.2, 46.0) | (219.55, −206.61, 37.17) (259.27, 58.05, 55.11) | 0.729 0.039 | 7.053 2.062 |

**Note.** We present the minimum and maximum linking length of each substructure, and the meaning of the other columns is the same as in Table 1.

However, this scatter could also be due to contamination from other substructures. To have a large sample of VMP stars in GSE, we adopted an eccentricity-based selection as this method is shown to have the highest completeness (Carrillo et al. 2024). While this sample has a lower purity, we do not consider the contamination to be significant. Two major sources of the contamination using the eccentricity criterion are the smooth components in the Galactic halo and the metal-weak tail of the kinematically heated high-$\alpha$ disk (Feuillet et al. 2020; Carrillo et al. 2024). The clustering algorithm we adopted should reduce the contamination from the smooth components, and the contamination of the high-$\alpha$ disk should not be significant at [Fe/H]< −2 since the number of high-$\alpha$ disk members drops sharply toward low metallicity. In addition, the distribution of our GSE stars in the $J_r$−$L_z$ plane ($\sqrt{J_r}$ = 25.5 kpc$^{1/2}$ km$^{1/2}$ s$^{-1/2}$ and $L_z$ = 113.8 kpc km s$^{-1}$) is similar to chemically selected GSE stars in Buder et al. (2022; $\sqrt{J_r}$ = 26 kpc$^{1/2}$ km$^{1/2}$ s$^{-1/2}$ and $L_z$ = 100 kpc km s$^{-1}$).

GSE members do not have a clear decreasing trend in $\alpha$-elements with metallicity, which is in agreement with the previous studies; that is, the [$\alpha$/Fe] ratio begins to decrease at [Fe/H] ⩾ −2.0 (e.g., −2.0 in Matsuno et al. 2019; −1.6 in Monty et al. 2020; −1.3 in Mackereth et al. 2019; and −1.1 in Horta et al. 2023). We notice that a large proportion of the Mg-poor stars ([Mg/Fe] < 0, as defined in Paper II) are associated with GSE, and this Mg-poor population in GSE is also found by Horta et al. (2023) at −1.8 < [Fe/H] < −0.8 based on data from APOGEE DR17. Among our HR sample, 50% (two out of four) of the Mg-poor stars belong to GSE, and this fraction is higher than the proportion of GSE to the whole sample (25%). These Mg-poor stars at all metallicity ranges could originate from dwarf galaxies accreted to GSE or from satellite dwarf galaxies of the GSE progenitor; see more detailed discussions in Section 4.2.2.

Previous studies have revealed the r-process enhancement of GSE at [Fe/H]> −2 using the GALAH Survey catalog (Matsuno et al. 2021; Myeong et al. 2022; da Silva & Smiljanic 2023) and elemental abundances derived from other high-resolution spectroscopic studies (Aguado et al. 2021a; Koch-Hansen et al. 2021; Carrillo et al. 2022; Naidu et al. 2022). Despite different selection methods to define GSE, those GSE samples all show enhancements of Eu and relatively low [Ba/Eu] ratios. In our sample, certain members of GSE are also r-process enhanced. The median Eu abundance of the GSE members with Eu measurements is 0.59, which is similar to the results from Aguado et al. (2021a; [Eu/Fe] = 0.59) and da Silva & Smiljanic (2023; [Eu/Fe] = 0.52). Specifically, GSE in our HR sample contains 16 r-process-enhanced stars, of which 11 are r-I stars[12] and five are r-II stars.[13] Two of the five r-II stars have also been associated with GSE by Y20.

However, when compared to our whole HR sample, GSE stars do not display higher Eu abundances and exhibit a scatter as large as the other sample stars. This may be due to the low metallicity of our sample since the large scatter of neutron-capture elements at [Fe/H]< −2 has been acknowledged for decades (e.g., McWilliam 1998; Johnson & Bolte 2002; Aoki et al. 2005; Sneden et al. 2008 and references therein) and is also clearly shown in Paper II. The scatter among halo stars can be explained by stochastic enrichments of r-process elements as a result of the hierarchical formation of galaxies and the inhomogeneity in the interstellar medium (Hirai et al. 2015; Ishimaru et al. 2015). It is plausible that the interstellar medium of the GSE was inhomogeneous at its early phase of evolution, i.e., at low metallicity, and GSE itself might be a product of mergers of smaller galaxies. We also note that our Eu abundance distribution might be biased because of the detection limit at low metallicity, where Eu lines cannot be detected if a star has an intrinsically low Eu abundance.

---

[12] The r-I stars are defined as those with 0.3 ⩽ [Eu/Fe] ⩽ 1.0, [Ba/Eu] < 0, following Sakari et al. (2018). This is slightly different from the definition we used in Paper II.
[13] [Eu/Fe] > 1.0, [Ba/Eu] < 0, which is the same as that used in Paper II.

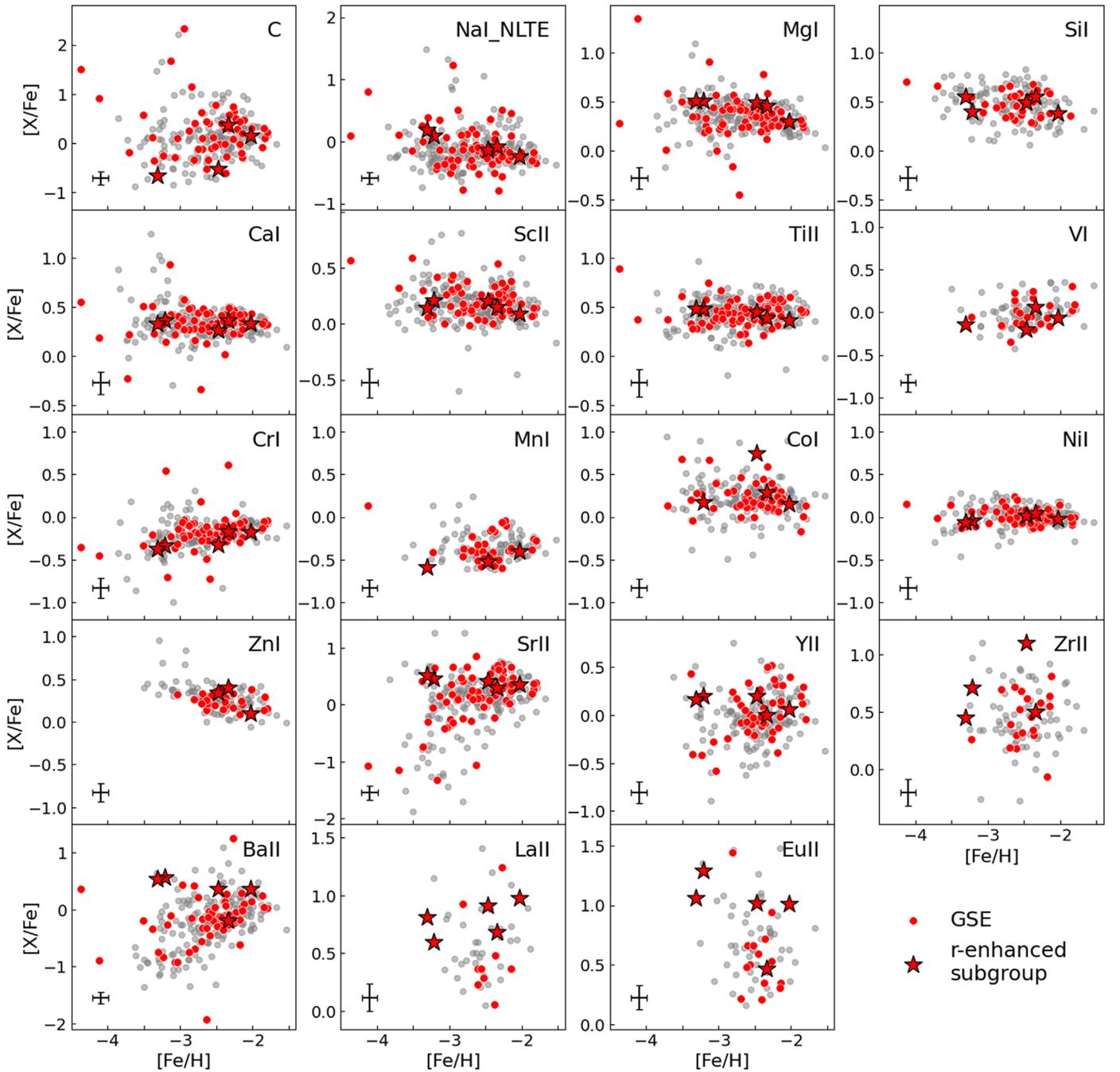

**Figure 7.** The chemical distribution of GSE. The element species is shown in the upper right of each subplot, and the average uncertainty is in the lower-left region. Red dots represent GSE member stars and gray dots represent all stars with reliable measurements. The red stars with black borders are members of the *r*-process-enhanced subgroup in GSE; see Section 4.2.3 for details.

#### 4.1.2. VMPD

A thick-disk-like component of metal-poor stars, which is known as the metal-weak thick disk, has been noticed for a long time and studied using various data sets (Norris et al. 1985; Morrison et al. 1990; Beers & Sommer-Larsen 1995; Chiba & Beers 2000; Beers et al. 2014; Li & Zhao 2017; Carollo et al. 2019; Naidu et al. 2020). Recently, as stellar samples expand toward low metallicities, VMP stars with disk-like kinematics have been gradually revealed (e.g., Sestito et al. 2019, 2020, 2021; Di Matteo et al. 2020; Venn et al. 2020; Carter et al. 2021; Cordoni et al. 2021; Mardini et al. 2022; Bellazzini et al. 2024; Carollo et al. 2023; Dovgal et al. 2024). However, the origin of these disk-like VMP stars is still debated. For example, using the NIHAO-UHD simulations (Wang et al. 2015; Buck 2020), Sestito et al. (2021) propose two scenarios to explain the origins of these stars: the dominant part of them have formed in the building blocks of the proto-Galaxy, and the remaining stars have originated from later merger events.

Several stars in our sample also demonstrate planar orbits with moderate eccentricity, providing us an opportunity to study the chemical properties of disk-like VMP stars. Based on the distribution in the $ec - |z|_{max}$ panel of our sample (Figure C1), we defined the disk-like VMP component as VMPD using $0.25 < ec < 0.60$ and $|z|_{max} < 3$ kpc. In addition, we also used rotational velocities and velocity distributions in the Toomre diagram to constrain VMPD with $v_\phi > -200$ km s$^{-1}$ and $(v_y - 233.1)^2 + v_x^2 + v_z^2 < 180$ km s$^{-1}$. The kinematic properties and metallicity distribution of our VMPD are similar to the low-eccentricity low-$|z|_{max}$ prograde VMP component as identified in, e.g., Cordoni et al. (2021).

The chemical distribution of VMPD is shown in Figure 8. VMPD displays a smaller scatter among most elements compared to the whole sample, except for J0705+2552, which is a CEMP star with a noticeable enhancement in Ba ([Ba/Fe] = 0.64) represented by the star symbol. Although this star does not meet our criterion of CEMP-s stars (see Section 4.2.1), the nonnegligible enhancement of $\alpha$-elements and s-process elements (e.g., C, Mg, Y, Ba) is similar to the chemical pattern of CEMP-s stars, which indicates that it could have been contaminated by an asymptotic giant branch (AGB) companion. We thus exclude this star from the following discussions on VMPD.

As shown in Figure 8, one clear feature of VMPD is its relatively low Zn abundance. The median [Zn/Fe] of VMPD is 0.16 dex, while that of the entire HR sample is 0.25 dex. We conduct a Kolmogorov–Smirnov (KS) test on the distribution of [Zn/Fe] at [Fe/H] $\geqslant -3$ between VMPD and the whole HR sample. The result shows $D = 0.47$ with $p = 0.02$, indicating a significant difference. One suggested main production site of Zn for VMP stars is hypernovae (HNe), whose explosion energy is one order of magnitude higher than regular supernovae (Umeda & Nomoto 2002; Kobayashi et al. 2006; Tominaga et al. 2007). In this scenario, a high [Zn/Fe] ratio would indicate either a high HN fraction or a high HN upper mass limit (e.g., Grimmett et al. 2020), and hence it would correspond to the production of Zn from relatively more massive stellar systems. For example, Mucciarelli et al. (2021) have reported that a globular cluster (GC) in the Large Magellanic Cloud (LMC), NGC2005, shows significantly lower [Zn/Fe] than the other GCs in the LMC or those in the Milky Way. They argued that the low Zn abundance could be explained by a chemical evolution model for a small dwarf spheroidal galaxy that has merged with the LMC a long time ago.

Following the same reasoning, the low Zn abundance of VMPD suggests that its progenitor systems might have been small and experienced slow star formation. Our result supports the first scenario by Sestito et al. (2021) that the disk-like low-metallicity stars could have originated from low-mass building blocks that merged in the very early Universe and initiated the formation of the Milky Way.

*4.1.3. Retrograde Substructures: Thamnos and Sequoia/I'itoi*

An excess of stars in the retrograde halo is considered to be one of the signatures of early accretion events (Helmi et al. 2017; Myeong et al. 2018a, 2018b). Helmi et al. (2018) proposed that these retrograde stars belong to Gaia-Enceladus. However, the differences in chemical properties suggest that the retrograde components and GSE may come from different merger events, which were subsequently identified as Sequoia and Thamnos, distinguished by their higher and lower energies (Koppelman et al. 2019a; Matsuno et al. 2019; Myeong et al. 2019). Naidu et al. (2020) argued that the high-energy retrograde halo should be further divided into three, Arjuna, Sequoia, and I'itoi, according to their different metallicity distributions, peaked at $-1.2$, $-1.6$, and $< -2$, respectively. In the following discussion, we still refer to this high-energy retrograde halo as Sequoia since the presence of the three subsubstructures has not been established yet. The chemical distributions of Thamnos and Sequoia are shown in Figure 9.

We assigned the largest number of retrograde stars to Thamnos. This is in agreement with a previous study by da Silva & Smiljanic (2023), whose most-retrograde groups, dominated by Thamnos, contain more VMP stars than other retrograde and prograde clusters. There is no obvious slope in Mg of Thamnos. Horta et al. (2023) showed that Thamnos presents no $\alpha$-knee feature at [Fe/H] $> -2$ using elemental abundances from APOGEE DR17, which suggests that Thamnos likely quenched star formation before the onset of Type Ia supernovae (SNe Ia). Our results do not contradict their conclusion.

Monty et al. (2020) found that for iron-peak elements (Mn, Zn) and neutron-capture elements (Y, Ba), the distribution of Thamnos has no obvious differences compared with other halo stars, which is also confirmed in our analysis. Da Silva & Smiljanic (2023) proposed that two [Eu/Mg] sequences are found in their Thamnos-dominated most-retrograde groups. One sequence has lower metallicity and [Eu/Mg] $\sim 0.4$, and the other has [Fe/H] $> -1.5$ and [Eu/Mg] $\sim 0.0$, which may be the contaminant from in situ components. All four Thamnos stars with Eu measurements in our sample show [Eu/Mg] $> 0.15$, and the average value is 0.31, which is similar to the low-metallicity r-process-enhanced sequence in da Silva & Smiljanic (2023).

There are six Sequoia stars with robust [Fe/H] measurements. Although low $\alpha$-element abundances (e.g., [Mg/Fe] $< 0.2$) have been reported for Sequoia at [Fe/H] $> -2$ (Matsuno et al. 2019; Monty et al. 2020; Matsuno et al. 2022b; Horta et al. 2023), our VMP Sequoia stars do not show such low Mg abundances. This indicates that there is a decreasing trend in these elemental abundances of Sequoia, supporting the idea that the low $\alpha$-element abundances at [Fe/H] $> -2$ are due to the large contribution from SNe Ia to the chemical enrichment in Sequoia. Unlike what was demonstrated in Matsuno et al. (2022b), the Na and Zn of Sequoia are not lower than those of other stars in our HR sample. In addition, the low Y abundance of Sequoia indicated by Matsuno et al. (2022b) is not seen in most of our Sequoia stars except for one star with [Y/Fe] $= -0.62$ and [Y/Mg] $= -0.99$. These differences are likely due to the different metallicity coverage; our sample has [Fe/H] $< -2$, while Matsuno et al. (2022b) covers $-1.8 <$ [Fe/H] $< -1.4$, where SNe Ia have started to contribute to the chemical enrichment of the progenitor of Sequoia.

There are four stars with robust Eu abundance measurements ($\sigma_{[Eu/Fe]} \leqslant 0.2$) in our six Sequoia stars, three of which are r-I stars. These fractions are higher than the proportion in the entire sample, where $\sim 22\%$ of stars have reliable Eu abundance measurements and $\sim 20\%$ of stars are r-process enhanced. This implies that Sequoia tends to have higher abundances of r-process elements on average. Aguado et al. (2021a) suggested that Sequoia members are r-process enhanced and show a tight and flat sequence with [Ba/Eu] $\sim -0.7$. Considering that Aguado et al. (2021a) selected Sequoia members with higher energy and that Koppelman et al. (2019a) proposed a more strict selection method of Sequoia with $E > -1.35 \times 10^5$ km$^2$ s$^{-2}$, we divided the Sequoia into two groups with higher ($E > -1.35 \times 10^5 \text{ km}^2 \text{ s}^{-2}$) and lower energy ($E < -1.35 \times 10^5 \text{ km}^2 \text{ s}^{-2}$) parts. All the three stars in the high-energy Sequoia are r-I stars with [Eu/Fe] ∼ 0.7 and higher Sr and Ba abundances than the entire sample of the present work, which is similar to the finding of Aguado et al. (2021a) but at lower metallicity.

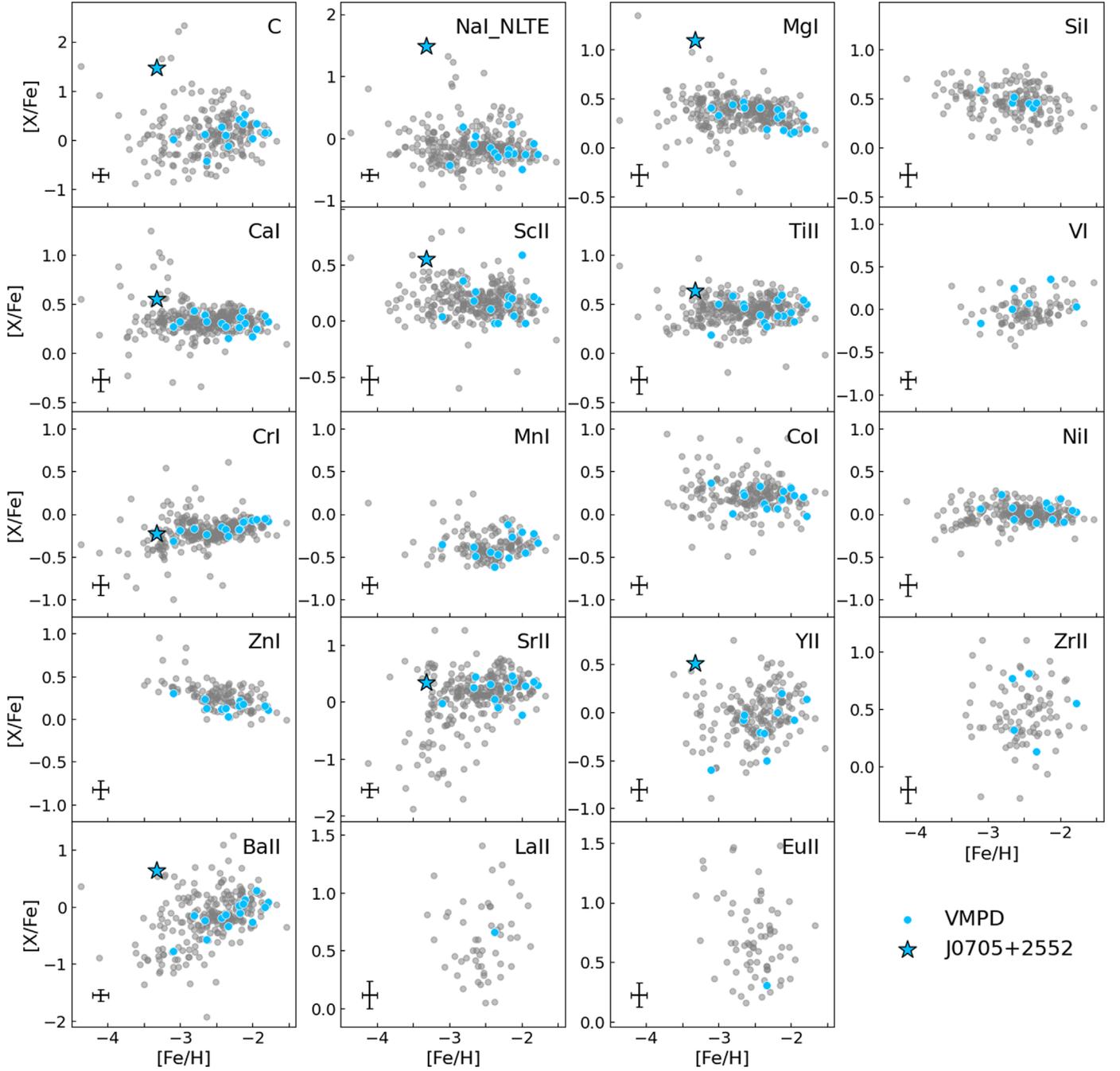

**Figure 8.** The chemical distribution of VMPD. Blue dots represent VMPD member stars, and the blue star with black border is J0705 + 2552, which is a CEMP star that might have been affected by mass transfer from an AGB companion. The meanings of other symbols are the same as in Figure 7.

### 4.1.4. Polar Substructures: Helmi Streams and Wukong/LMS-1

Helmi streams are characterized by their large perpendicular component of angular momentum and the polar ring-like orbits of their members in the solar vicinity (Helmi et al. 1999; Koppelman et al. 2018, 2019b). Another polar substructure has been identified after Gaia releases by Yuan et al. (2020a) and Naidu et al. (2020), LMS-1 and Wukong, respectively. Due to their similar dynamics and the same GC memberships, LMS-1 and Wukong are considered to be the same substructure. We will refer to this substructure as Wukong in the following discussion. The chemical distributions of Helmi streams and Wukong are shown in Figure 10.

We do not find CEMP stars in Helmi streams, and all stars with reliable carbon measurements ($\sigma_{\text{[C/Fe]}} \leqslant 0.2$) have relatively low carbon abundances with [C/Fe] ⩽ 0.11. This fraction of CEMP stars in Helmi streams is significantly lower than the previous estimates (∼20%; see Aguado et al. 2021b; Gull et al. 2021). One possible reason for this result is that members of Helmi streams in our sample are mainly giants (11/15), while in Aguado et al. (2021b) they are mainly main-sequence/turnoff stars (54/62).[14] As discussed in Section 3.4 of Paper II, the fraction of CEMP stars is higher in turnoff stars (22%) than in giants (7.8%), so a higher giant fraction may indicate a lower fraction of CEMP stars. However, most of the stars in Gull et al. (2021) are also giants (10/12), and two CEMP stars are found. The fraction of CEMP stars will be discussed in detail in Section 4.2.1.

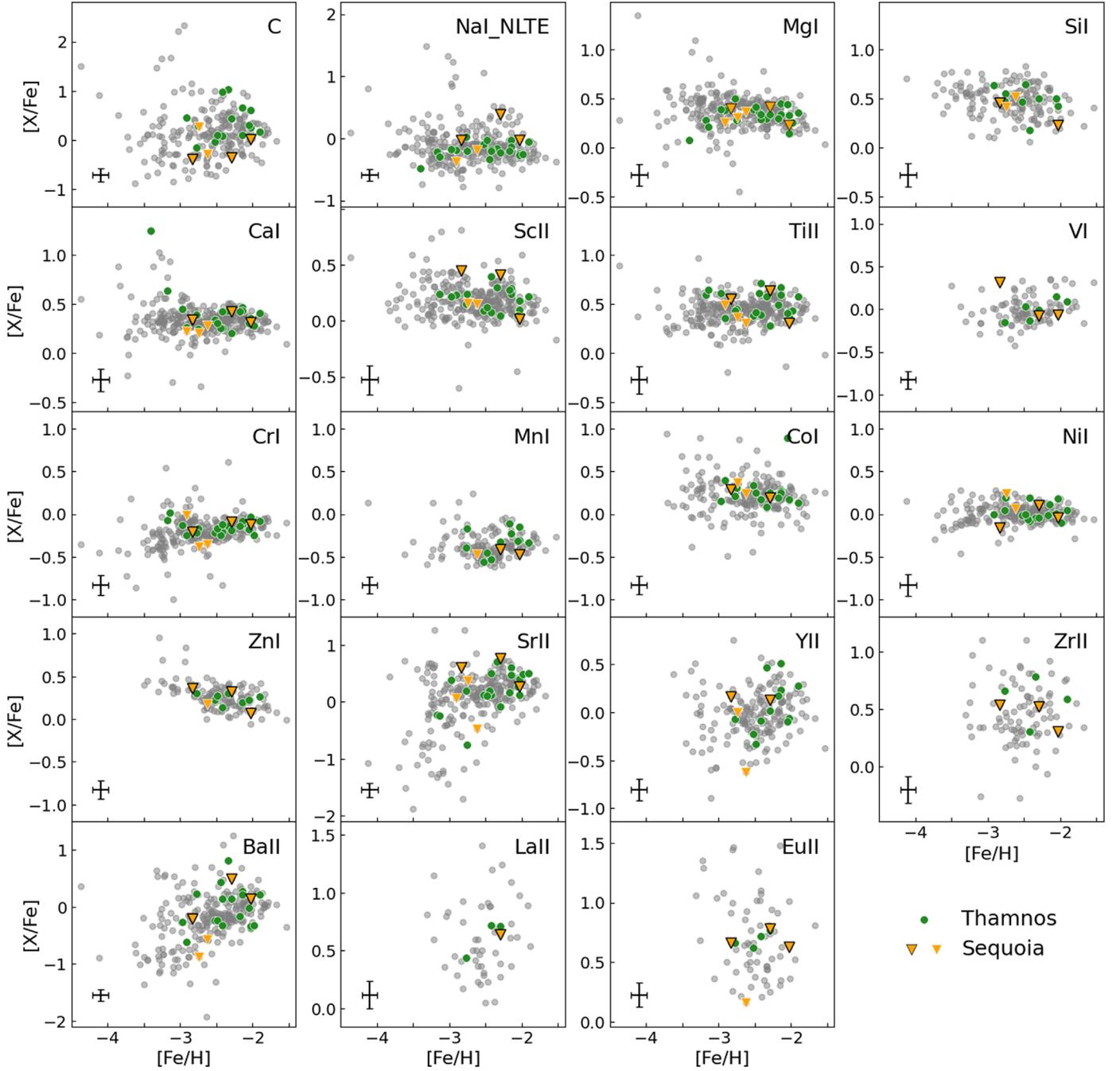

**Figure 9.** The chemical distribution of Thamnos and Sequoia. Colored symbols represent member stars in Thamnos (green dots) and Sequoia (orange inverted triangles). The Sequoia members are divided into two parts, presented by inverted triangles with black borders ($E > -1.35 \times 10^5$ km$^2$ s$^{-2}$) and with white borders ($E < -1.35 \times 10^5$ km$^2$ s$^{-2}$). The meanings of other symbols are the same as in Figure 7.

The low CEMP fraction based on our HR sample does not contradict the previous findings that Helmi streams have a relatively low C abundance. For example, Roederer et al. (2010) reports subsolar carbon abundances of Helmi stream stars, and Gull et al. (2021) also show that the fraction of CEMP stars is lower in Helmi streams than in other dwarf galaxies such as Sculptor and the progenitor of $\omega$ Cen, which agrees with our result.

Previous studies have found low Mg abundances of Helmi streams at [Fe/H] > −2.0 (e.g., [Mg/Fe] ∼ 0.2 in Aguado et al. 2021b and [Mg/Fe] ∼ 0.1 in Matsuno et al. 2022a), while most of our Helmi stream members have [Fe/H]< −2.0 with a median Mg abundance of [Mg/Fe] = 0.4. This implies that Helmi streams should have a decreasing trend in Mg abundance as found in other substructures (e.g., GSE and Sequoia).

---

[14] We use same criteria as in Paper II to define turnoff stars ($T_{\rm eff} \geqslant 5500$ K & log $g \geqslant 3.0$), red giant branch (RGB) stars ($T_{\rm eff} < 5500$ K), and horizontal branch (HB) stars ($T_{\rm eff} \geqslant 5500$ K & log $g < 3.0$). Both RGB stars and HB stars are considered giants in this work.

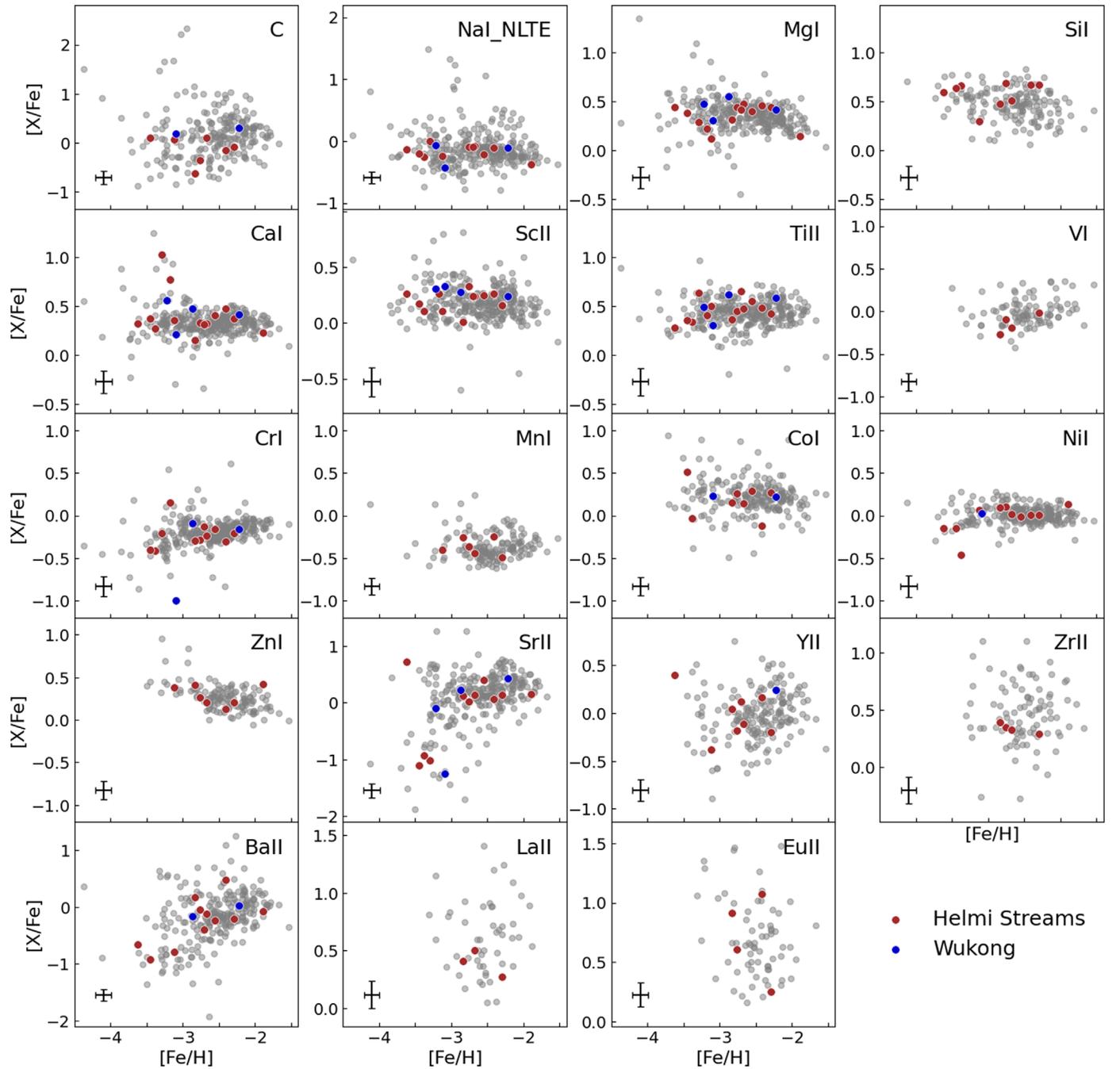

**Figure 10.** The chemical distribution of Helmi streams and Wukong/LMS-1. Colored dots represent member stars in Helmi streams (brown) and Wukong/LMS-1 (blue). The meanings of other symbols are the same as in Figure 7.

The members of Helmi streams show lower light neutron-capture element abundances, such as Sr and Zr, compared with the entire sample. For [Fe/H] > −3, where the scatter of Sr is relatively small, the median Sr abundance of Helmi streams is 0.14, which is ∼0.1 dex lower than the whole sample, and for [Fe/H] < −3 three Helmi stream members have extremely low Sr abundances with [Sr/Fe] ∼ −1. The median Zr abundance of Helmi streams is also ∼0.1 dex lower than the entire sample, while another light neutron-capture element, Y, does not have such a feature. Although low light neutron-capture element abundances of Helmi streams are also noticed by Aguado et al. (2021b) and Matsuno et al. (2022a), they report a more significant deficiency of Sr and Y in the Helmi stream (>0.5 dex) than our findings.

Unlike the light neutron-capture elements, the distribution of heavier neutron-capture elements does not show significant differences between Helmi streams and the other stars. Among the four stars with robust Eu measurements in Helmi streams, three are *r*-process enhanced, of which two are *r*-I stars and one is an *r*-II star. This indicates that the *r*-process dominates the nucleosynthesis of the heavier neutron-capture elements in Helmi streams. This is consistent with previous studies (Limberg et al. 2021b; Gull et al. 2021).

The deficiency of light neutron-capture elements in Helmi streams can be attributed to the lack of production sites of these elements in the early Universe within its progenitor, such as rotating massive stars (Frischknecht et al. 2012; Choplin et al. 2018), collapsars (Siegel et al. 2019), and electron-capture supernovae (Wanajo et al. 2011). However, the distribution of heavier neutron-capture elements in Helmi streams is similar to that of other stars, hence the progenitor of Helmi streams likely lacked rotating massive stars, which was also proposed by Matsuno et al. (2022a). Rotating massive stars might also contribute to the formation of some CEMP-no stars at extremely low metallicity (e.g., see Maeder & Meynet 2015; Maeder et al. 2015; Liu et al. 2021; Jeena et al. 2023, and references therein).[15] A smaller number of rotating massive stars also explains the lower carbon abundances of Helmi streams to some extent.

We associated DTGs with Helmi streams using the *z*-component and the perpendicular component of angular momentum, following the criterion proposed by Koppelman et al. (2019b). The member stars in Helmi streams show two sequences in radial action phase space, which is also noticed by Y20 and studied in detail by Dodd et al. (2022). Compared with the low radial action component, stars with higher radial action have higher energy but the same vertical motion and action. As discussed in Y20, this high radial action part may be stripped at an earlier pericentric passage from the progenitor of Helmi streams or come from a different origin. Matsuno et al. (2022a) suggested that the stars with high energy ($E \sim -1.0 \times 10^5 \, \mathrm{km^2 \, s^{-2}}$) may not come from the same progenitor as Helmi streams, because the high-energy star Gaia EDR3 2447968154259005952 has different chemical abundance ratios in $\alpha$-elements and neutron-capture elements from the other stars in their sample. However, in our HR sample, there are no significant differences in the chemical distribution between the high radial action sequence and the other members of Helmi streams, so both of these two sequences are considered Helmi streams.

The chemodynamical studies of the Wukong substructure are still limited, and its selection criteria are still unclear. In this work, we refer to a relatively strict criterion from Limberg et al. (2024) to associate stars in our sample with Wukong. We have only four Wukong members from the HR sample.

Wukong stars have relatively higher $\alpha$-element abundances and do not show a significant decreasing trend at [Fe/H] $< -2.2$, which is consistent with results from Limberg et al. (2024), indicating a relatively massive progenitor of Wukong. For other elements, there are too few stars in Wukong to discuss their chemical properties, so we only provide the distribution of these stars for reference.

### 4.1.5. Other DTGs

As discussed in Section 3.2 and presented in Figure 2, there are four DTGs, namely DTG-8, DTG-9, DTG-20, and DTG-49, which contain at least three HR sample stars but are not associated with known substructures. Recently, Malhan (2022) identified a new substructure named Pontus in the retrograde halo using the clustering algorithm, which overlaps with GSE and Thamnos in the energy–action phase space. Two DTGs in our sample might be associated with Pontus. Given that the chemical and dynamical properties of Pontus remain unclear, we discuss it together with new DTGs in this paper. The chemical distributions of these four DTGs and Pontus are shown in Figure 11.

DTG-8 is a prograde group with low-eccentricity orbits, but it is not associated with VMPD because of its relatively large $|z|_{\max}$. DTG-8 also shows different chemical distributions from VMPD. Specifically, the members of DTG-8 show heavy element enhancements and no significant deficiency in Zn. Two stars in DTG-8 with Eu abundance measurements are both *r*-II stars, and the [Ba/Eu] of these two stars are $-0.59$ and $-0.80$, implying an *r*-process-dominated nucleosynthesis for heavy elements in DTG-8. The [Eu/Mg] of these two stars are 0.63 and 1.08, which requires a significant *r*-process element production in the progenitor system of DTG-8 through several *r*-process events other than core-collapse supernovae (CCSNe), such as neutron star mergers. Although the other five stars do not have Eu measurements, the enhancement of other heavy elements, such as Ba and La, still indicates that DTG-8 is likely enhanced in heavy elements in general.

DTG-9 is a polar component with small net rotational velocity, which has been noticed by several works (e.g., Y20; Limberg et al. 2021a; Shank et al. 2022a) and labeled as ZY20:DTG-35, GL21:DTG-8, and DS22:DTG-6, respectively.[16] DTG-20 is a planar substructure with moderate eccentricity, whose median $|z|_{\max}$ is lower than 1 kpc. The chemical distributions of DTG-9 and DTG-20 are similar to those of the entire sample. However, the number of stars in these two groups is not large enough to draw any conclusions about their progenitors.

DTG-49 is a high-energy prograde group that has also been clustered by Y20 as ZY20:DTG-2. DTG-49 shows a significant deficiency in Na and Mg, and two of the stars also have low Ca and Ti abundances, suggesting a low star formation efficiency in the progenitor of this group. Only three stars are assigned to this group in the HR sample, so it is difficult to obtain any reliable constraints on its progenitor. However, such low Mg and Na abundances (e.g., [Mg/Fe] = 0.05 and [Na/Fe] = $-0.51$) at [Fe/H] $\sim -3$ imply that it is likely to be from an accreted low-mass dwarf galaxy, which provides us with a unique perspective to investigate the chemical properties of low-mass dwarf galaxies with solar neighborhood stars. Therefore, a high-resolution follow-up of other DTG-49 members in the LR sample is highly desired.

Pontus has three stars with elemental abundances, one of which has a metallicity uncertainty greater than 0.2 and should be excluded from the chemical discussion. Therefore, only two Pontus members are available, which is too few to draw any conclusions from. We simply provide the chemical distribution of these two stars in Figure 11 for reference. We notice that only two stars are associated with Pontus in Horta et al. (2023), too. A larger sample of Pontus is needed for the further study of its chemical properties and its relationship with GSE and Thamnos in detail.

---

[15] CEMP-no stars are CEMP stars that do not present enhancements in heavy elements.

[16] Here, we follow the nomenclature proposed by Y20 and use XXYY:DTG-x to refer to DTGs identified in previous work, where XX are the initials of the names of the first author, YY are the last two digits of the year of publication, and DTG-x is the number of the DTG. For example, our DTG-1 will be referenced as RZ24:DTG-1.

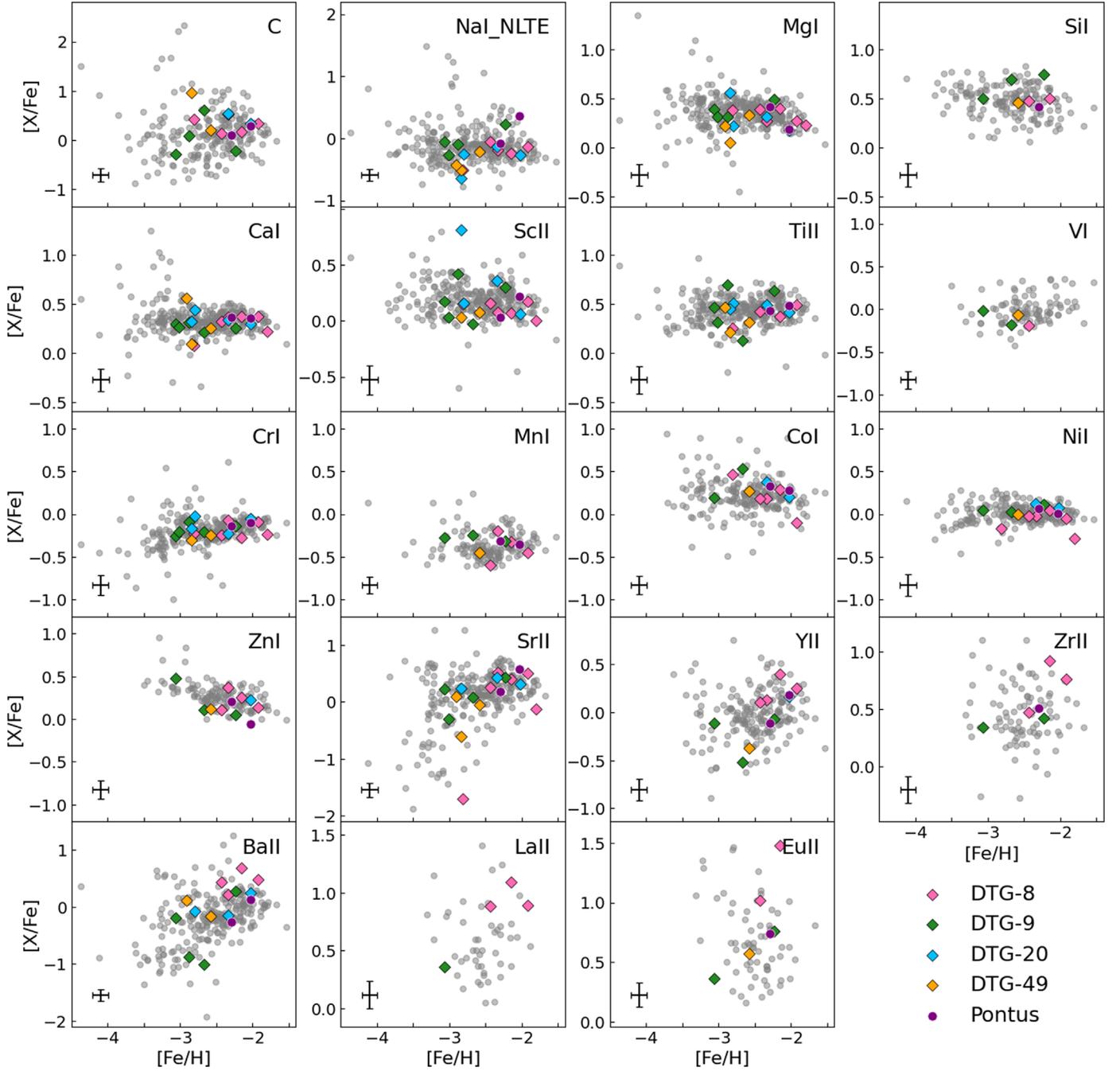

**Figure 11.** The chemical distribution of DTG-8, DTG-9, DTG-20, DTG-49, and Pontus. Colored diamonds represent member stars in DTG-8 (pink), DTG-9 (green), DTG-20 (blue), DTG-49 (orange), and Pontus (purple). The meanings of other symbols are the same as in Figure 7.

### 4.2. Kinematics of Stars with Particular Abundance Features

The chemical properties within each individual substructure and DTG have been discussed in Section 4.1. In the following section, we will compare the kinematic properties of CEMP stars, $\alpha$-peculiar stars, and neutron-capture-element peculiar stars. Note that we do not exclude stars with relatively large abundance uncertainties in this section, and thus the number of stars having these particular abundance features is slightly different from those in Section 4.1.

#### 4.2.1. CEMP Stars

The distribution of CEMP (especially CEMP-no) stars could be rather sensitive to the environments of their progenitor systems. It is thus of paramount interest to investigate the fraction and the C abundance distribution of CEMP stars in our HR sample.

Adopting the same criterion of CEMP stars as in Paper II, i.e., [C/Fe] $\geqslant +0.7$ for stars with $\log(L/L_\odot) \leqslant 2.3$ and [C/Fe] $\geqslant 3.0 - \log(L/L_\odot)$ for more luminous objects (Aoki et al. 2007), there are 47 CEMP stars in the HR sample. Note that adopting the correction on the stellar evolution by Placco et al. (2014) would not make any difference in the selected CEMP

Table 3
CEMP Star Fraction of Known Substructures, DTGs, and Nonclustered Stars

| Substructure | CEMP | CEMP-s | | CEMP-no | | C-normal[a] | | Members | | CEMP-no ratio[b] | |
|---|---|---|---|---|---|---|---|---|---|---|---|
| | | Turnoff | Giants | Turnoff | Giants | Turnoff | Giants | Turnoff | Giants | Turnoff | Giants |
| GSE | 12 | 2 | 0 | 7 | 3 | 20 | 35 | 46 | 43 | 25.9% | 7.9% |
| VMPD | 2 | 1 | 1 | 0 | 0 | 8 | 7 | 10 | 8 | 0 | 0 |
| Thamnos | 4 | 1 | 0 | 3 | 0 | 7 | 8 | 15 | 8 | 30.0% | 0 |
| Helmi streams | 0 | 0 | 0 | 0 | 0 | 2 | 8 | 4 | 11 | 0 | 0 |
| Sequoia | 1 | 0 | 0 | 0 | 1 | 0 | 5 | 1 | 6 | 0 | 16.7% |
| Wukong | 0 | 0 | 0 | 0 | 0 | 1 | 1 | 2 | 2 | 0 | 0 |
| Pontus | 1 | 0 | 0 | 1 | 0 | 1 | 1 | 2 | 1 | 50.0% | 0 |
| other DTGs | 9 | 1 | 1 | 6 | 1 | 12 | 38 | 27 | 41 | 33.3% | 2.6% |
| Nonclustered | 19 | 3 | 2 | 9 | 5 | 14 | 61 | 47 | 78 | 39.1% | 7.6% |
| Total | 48 | 8 | 4 | 26 | 10 | 65 | 164 | 154 | 198 | 28.6% | 5.7% |

**Notes.**
[a] C-normal refers to stars with a [C/Fe] value or upper limit less than 0.7.
[b] CEMP-no ratio = CEMP-no/(CEMP-no + C-normal), if the numbers of CEMP-no and C-normal stars are both equal to 0, this ratio is 0.

stars from the HR sample, as pointed out by Paper II. As discussed at the beginning of Section 4, ten of them with [Ba/Fe] > 1.0 are classified as CEMP-s stars. Additionally, based on noticeable enhancement in Ba and other s-process elements, another CEMP star J0705+2552 (discussed in Section 4.1.2) and a mildly C-enhanced ([C/Fe] = 0.56) giant J0446+2124 are also treated as (potential) CEMP-s stars in this study. In summary, there are 12 CEMP-s stars and 36 CEMP-no stars.

To investigate whether the fraction of CEMP-no stars varies with their birth environments, we examined numbers and ratios of CEMP-no stars in different substructures, as shown in Table 3. For DTGs not associated with known substructures, including the four large DTGs discussed in Section 4.1.5, they are referred to as other DTGs. All of the nonclustered stars are considered as a mixture of smooth components and small accreted components that have too few member stars to be clustered in our sample. The CEMP-no ratios are computed by CEMP-no/(CEMP-no + C-normal), where C-normal refers to stars with a [C/Fe] value or upper limit less than 0.7. In general, for the clustered stars, the fraction of CEMP-no stars tends to be higher among turnoff stars than that among giants, which is consistent with the whole HR sample. For GSE and Thamnos, the fractions of CEMP-no stars in both turnoff stars and giants are similar to those in nonclustered stars. However, for VMPD and the Helmi streams, which have statistically meaningful numbers of objects in the HR sample, there are no CEMP-no stars detected. This may be related to the properties of their progenitor systems. For example, as shown in the merger-tree model by Salvadori & Skúladóttir (2015), CEMP-no stars should exist in all dwarf galaxies within the observed luminosity range; however, as the galaxy luminosity increases, the overall probability of observing CEMP-no stars decreases. Such a theoretical prediction has been recently confirmed by the discovery of an extremely metal-poor CEMP-no star in Sculptor (Skúladóttir et al. 2024). Therefore, a larger sample of VMPD and Helmi stream stars is required to explore whether the negative detection of CEMP-no stars in these two substructures reflects the properties of their progenitor systems or is caused by the limited size of our HR sample.

Yoon et al. (2016) classifies CEMP stars into three groups in the A(C)−[Fe/H] diagram that would reflect the primary sources of their carbon enhancements. We placed all CEMP stars in our sample in the A(C)−[Fe/H] diagram, as shown in Figure 12. There are quite a number of CEMP stars belonging to Group II and a few to Group III. For both groups, there are more clustered CEMP stars than the nonclustered ones, which is consistent with the hypothesis that these two groups are more closely related to dwarf and/or UFD galaxies (e.g., Yoon et al. 2019).

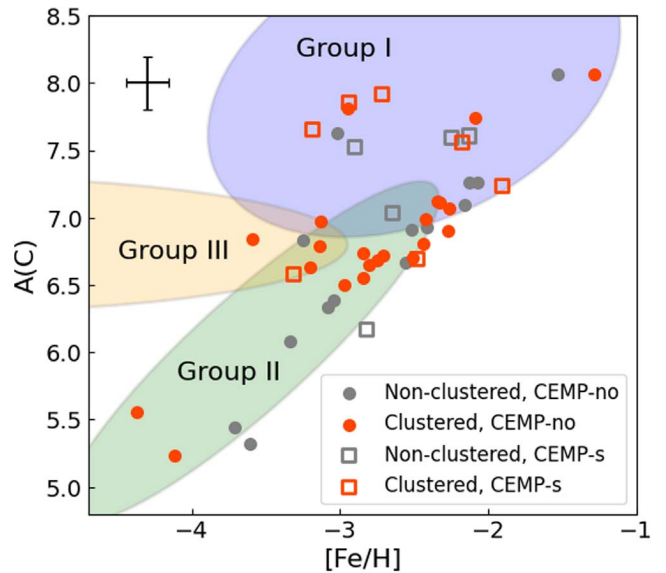

**Figure 12.** The C abundance distribution of all CEMP stars in the HR sample. The solid dots refer to CEMP-no stars and open squares refer to CEMP-s stars. Red and gray symbols represent the clustered and nonclustered CEMP stars, respectively. The average uncertainties are shown in the upper left of this figure. The morphological classification from the Yoon–Beers diagram (see Figure 1 in Yoon et al. 2016) has been overplotted.

Zepeda et al. (2023), analyzing CEMP stars from various abundance studies, identified 99 CEMP stars in GSE, 15 CEMP stars each in both Wukong and Thamnos, and none in the Helmi streams. The relative number of CEMP stars found in their compiled sample is in general agreement with our results; that is, compared with GSE, Wukong, and Thamnos, Helmi streams tend to contain fewer CEMP stars. Furthermore, the distribution of their sample in the A(C)−[Fe/H] diagram favors a scenario in which the Group II CEMP stars are more likely to be formed in low-mass dwarf galaxies that do not support extended star formation environments, which is consistent with what we have found based on our VMP star sample.

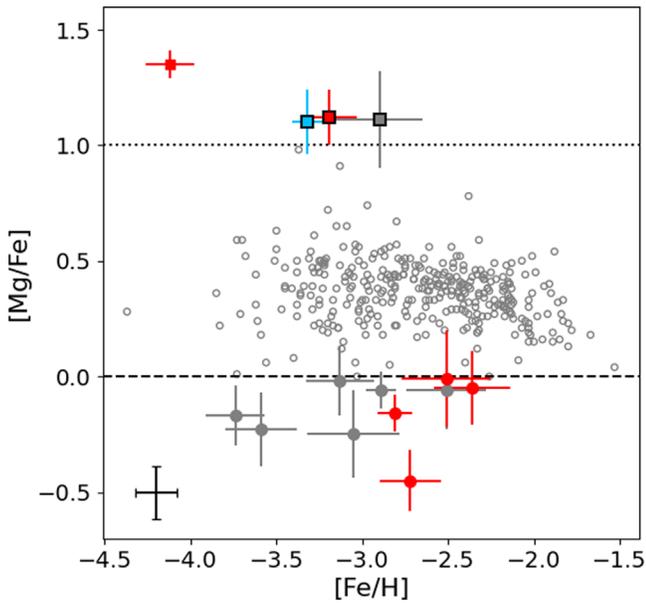

**Figure 13.** The Mg abundance distribution of the HR sample. All Mg-normal stars are represented by gray open dots and the average uncertainty of the Mg-normal stars is shown in the lower left of this panel. Mg-poor stars are represented by solid dots and Mg-rich stars by squares. The symbols with black borders stand for CEMP-s stars. Red, sky blue, and gray stand for members of GSE, VMPD, and nonclustered stars, respectively. The dashed and dotted lines are [Mg/Fe] = 0 and [Mg/Fe] = 1.

Moreover, we note that there is a peculiar ultra metal-poor CEMP-no star J2217+2104 ([Fe/H] = −4.12, A(C) = 5.23) showing [C/Mg]∼ −0.5, which has been studied in detail by Aoki et al. (2018). Based on the dynamical analysis in this work, J2217+2104 turns out to be an accreted star, which is associated with GSE. We also note that one CEMP-no star with [C/Fe] > 2, which is found in the area of Group I in Figure 12, is J0150+2149 ([Fe/H] = −2.95, A(C) = 7.81; Paper II). This star is also associated with GSE.

#### 4.2.2. α-peculiar Stars

We follow the criteria in Paper II to divide the HR sample stars into three parts according to their Mg abundances, namely Mg-poor stars ([Mg/Fe] < 0), Mg-rich stars ([Mg/Fe] > 1) and Mg-normal stars (0 ⩽ [Mg/Fe] ⩽ 1). As shown in Figure 13, there are ten Mg-poor stars and four Mg-rich stars in the HR sample.

Among the four Mg-rich stars, two are from GSE, one is from VMPD, and one is not clustered. All of the Mg-rich stars are CEMP stars. Three of them are CEMP-s stars whose abundances are influenced by their AGB companion, and the remaining one is the ultra metal-poor CEMP-no star J2217+2104, which is mentioned in Section 4.2.1. We do not discuss the kinematic properties of these Mg-rich stars because there is only one star if the CEMP-s stars are excluded.

Four Mg-poor stars are from GSE, and the remaining six are not clustered. Nonclustered stars may form in situ or come from dynamically relaxed accretion events. The deficiency in Mg can be explained by the contribution of SNe Ia (Ivans et al. 2003; Caffau et al. 2013) or pair-instability supernovae (PISNe, Aoki et al. 2014; Xing et al. 2023). SNe Ia can produce Ca but hardly produce Mg; therefore, a subsolar [Mg/Ca] ratio may indicate that the SNe Ia have contributed to the chemical enrichment of the interstellar matter, while PISNe would also lead to a subsolar [Mg/Ca], but lower [C/Fe] and [Na/Fe] are also expected. As discussed in Paper II, most of the Mg-poor stars show subsolar [Mg/Ca] ratios, but no Mg-poor stars fit the chemical pattern of PISNe in our sample, which is consistent with the increased contribution of SNe Ia. Therefore, these Mg-poor stars may have been born in dwarf galaxies with extremely low star formation rates, where the SNe Ia began to contribute to the chemical enrichment below [Fe/H]< −2.5. These dwarf galaxies were then accreted by a massive system, such as the Milky Way or GSE.

#### 4.2.3. Neutron-capture-element Peculiar Stars

As discussed in Sections 4.1.1 through 4.1.5, abundances of neutron-capture elements generally present rather large scatter in most substructures. Based on the chemodynamical analysis of our HR sample, there is no clear evidence indicating whether stars that have been accreted are more likely to be enhanced in neutron-capture elements. This may indicate that neutron-capture elements were synthesized by various production sites in the early times, which led to the diversity in the level of heavy elements enrichment in the low-metallicity region. However, there is one interesting dynamical subgroup of r-process-enhanced stars in GSE.

Among our GSE stars five are on particularly eccentric orbits, having high radial actions. Because of the slight differences in the vertical actions, they do not belong to a single DTG and, instead, belong to three DTGs. Interestingly, this dynamically defined group contains a high fraction of r-II stars (see stellar symbols in Figure 7), with four out of five stars being r-II stars; the remaining one is still an r-I star. All the five stars are giants with low [Ba/Eu] (Figure 14), indicating that their neutron-capture elements are mostly produced through the r-process nucleosynthesis. Moreover, they display similar abundance patterns in α-elements, iron-peak elements (e.g., Cr, Mn, Ni), and neutron-capture elements, suggesting that the five stars might share a common origin. These stars show a wide metallicity distribution from ∼ −3.5 to ∼ −2.0, which differs greatly from GCs.

We suggest that this r-process-enhanced subgroup of GSE may have originated from a dwarf galaxy similar to the UFD galaxy Ret II (Ji et al. 2016). It is not clear at the moment whether the progenitor galaxy has been accreted before or after the merger between GSE and the Milky Way or happens to have similar kinematics as GSE. A detailed study of this subgroup with a larger sample and theoretical simulations of galaxy mergers and chemical enrichments would be necessary to completely unravel its origin. It would also be of interest to study the abundance patterns for more elements, especially actinides.

This group is different from what has been studied in Roederer et al. (2018) and Hattori et al. (2023) in the sense that while they studied clustering in kinematics among r-rich stars, we found a group of stars clustering in kinematics that turned out to contain a high fraction of r-rich stars. It is thus the first example of a compact dynamical group with extreme r-process enhancement in the GSE. The stars are relatively bright (Gaia G < 15 mag) and nearby, providing us with a valuable laboratory to obtain full r-process element patterns of r-rich stars that were likely born in a dwarf galaxy. Further studies on a more complete abundance pattern of heavy elements, e.g., from Sr through Th, will enable us to reveal the origin of the r-process elements in this subgroup. Such detailed abundance patterns would also allow us to confirm the common origin of these stars if they turn out to be coherent within this group.

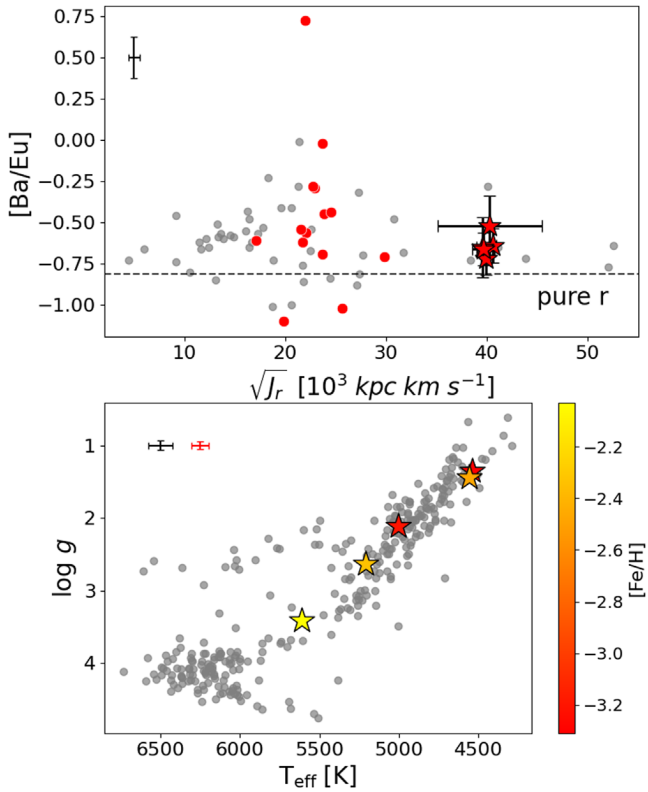

**Figure 14.** The distribution of the *r*-process-enhanced subgroup in GSE in the $\sqrt{J_r}$–[Ba/Eu] panel (top) and the HR diagram (bottom). The gray dots and the black error bars in the upper left of each panel are the value and typical uncertainties of the whole sample (median uncertainty for action and average uncertainties for other quantities), and the red dots in the top panel are all GSE member stars. The subgroup members are presented by stars colored in red (top) and colored by their metallicity (bottom). In the top panel, the pure-*r* production is from Frebel & Ji (2023). In the bottom panel, the red error bars represent the average uncertainties of the subgroup members.

We also study the dynamical origin of 25 neutron-capture-element-poor VMP stars whose [Sr/Fe] < −0.5 and [Ba/Fe] < −0.5. Twelve of these stars are not clustered in our sample, while five are not associated with the known substructures. Among the remaining eight stars, four are associated with GSE, two with Helmi streams, and one each from Thamnos and Wukong. There is no significant connection between the distribution of neutron-capture-element-poor VMP stars and the different substructures.

### 4.3. Global View

Here, we discuss the global trend and correlation among chemodynamics of halo stars without classifying stars into substructures or DTGs. Such global trends could potentially enable us to observe signatures of accreted galaxies that are yet to be discovered from kinematics. For example, Venn et al. (2004) found that there is a significant trend between α-element abundances and velocities of halo stars in the solar neighborhood in the sense that stars on extremely retrograde orbits have low [α/Fe] ratios. Thanks to recent advances in the field, the extremely retrograde component is now considered to correspond to a kinematic substructure, Sequoia (Matsuno et al. 2019; Monty et al. 2020). In addition, global trends might also provide constraints on the role of the halo mass of the accreted dwarf galaxies in their chemical enrichment and tidal interaction with the Milky Way. Since massive galaxies suffer from stronger tidal friction during the accretion onto the Milky Way, they are expected to deposit stars deeper into the gravitational potential of the Milky Way (Amorisco 2017). Therefore, if the chemical evolution of dwarf galaxies primarily depends on their halo masses, we might be able to observe correlations between chemical abundances and kinematics among halo stars in the present Milky Way. While, for example, Gratton et al. (2003), Venn et al. (2004), Ishigaki et al. (2012), and Monty et al. (2020) studied chemodynamical correlations among halo stars, including the very metal-poor range, we can now study such correlations among a large sample of VMP stars using precise kinematics from the Gaia mission.

We examine correlations between chemical abundance ratios and kinematics in Figures 15–17. We choose $J_\phi$, $\sqrt{J_r}$, and $\sqrt{J_z}$ as the variables representing kinematics of stars to maintain the consistency with our clustering analysis described earlier. The three quantities respectively enable us to investigate how the chemical properties vary with the orbital angular momentum in the disk rotation direction of the Milky Way, the radial excursion, and the vertical excursion. More massive accreted galaxies are expected to show small absolute values in the actions. We include in each panel Pearson's correlation coefficient and partial correlation coefficient, and the corresponding *p*-values with $p < 0.01$ are considered to be significant.[17] The partial correlations are adopted to minimize the combined effect of our sample selection and chemical evolution. For the panels showing correlations between [Fe/H] and kinematics, we take the distances to the stars as the controlling variable to test if the observed correlations can be attributed to the correlation between [Fe/H] and distance (Figure 2). For the other panels, we take [Fe/H] as the controlling variable since [Fe/H] may show correlations with kinematics due to the sample selection and because [X/Fe] ratios evolve with metallicity due to chemical evolution (see Figures 15, 18, and 20 in Paper II), which can then produce an artificial correlation between [X/Fe] and kinematics even when [X/Fe] evolves similarly with metallicity regardless of kinematics.

We observe a significant [Fe/H] gradient as a function of $\sqrt{J_z}$ in the sense that lower-metallicity stars tend to show large vertical actions. However, if we take the distance as the control variable and compute the partial correlation coefficient between [Fe/H] and $\sqrt{J_z}$, the correlation becomes insignificant. Hence, it is necessary to conduct a study of the metallicity gradient using a volume-complete sample that is unbiased in metallicity, such as in Youakim et al. (2020).

Regardless of the origin of the metallicity gradients, we can still determine if there are intrinsic correlations between [X/Fe] and kinematics since our sample selection does not bias [X/Fe] at a given metallicity. Figures 15–17 show that there are significant correlations between C, Na, V, Cr, Zn, Sr, and Ba abundances and $\sqrt{J_z}$, and between Sr abundance and $J_\phi$. Once conditioned on [Fe/H], all but the one between [C/Fe] and $\sqrt{J_z}$ become insignificant. Therefore, most of the significant correlations observed can be explained by the combination of the [Fe/H] gradient and the evolution of [X/Fe] with metallicity. Even though there is a metallicity gradient in the halo as a function of kinematics, abundance ratios at a fixed [Fe/H] do not seem to show gradients, indicating that there is not much variation in chemical evolution in the very metal-poor regime.

---

[17] We compute the *p*-values for partial correlation (*r*) by converting *r* to $t = r\sqrt{(N-3)/(1-r^2)}$, which follows the *t*-distribution with the $N - 3$ degree of freedom, where *N* is the number of data points.

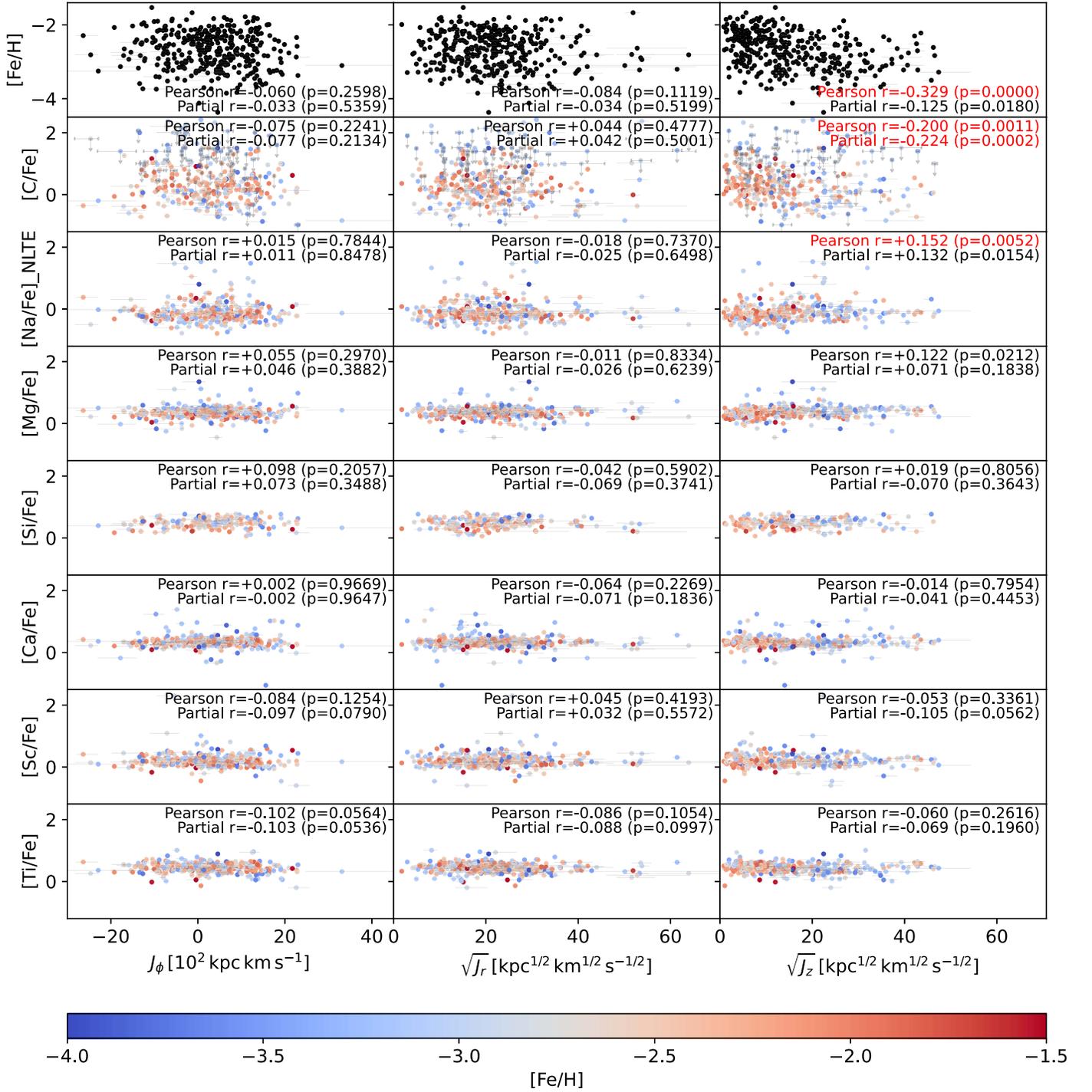

**Figure 15.** Correlations between chemical abundance ratios and actions. Except for the panels in the first row, we color code stars according to the metallicity. Partial correlation coefficients are computed taking the distance and [Fe/H] as the controlling variable for [Fe/H] and the other panels, respectively. We highlight correlations that are highly significant with the $p$-value smaller than 0.01.

The negative correlation between [C/Fe] and $\sqrt{J_z}$ needs attention. The correlation seems highly significant: the $p$-value is 0.0002, which is clearly below our threshold of $p = 0.01$. While we have 60 combinations of chemical abundance and kinematics in the figures, the probability of all 60 pairs of random variables showing $p > 0.0002$ is $1 - (1 - 0.0002)^{60} = 98.8\%$. This might be related to the varying fraction of CEMP stars in different substructures as we discussed in Section 4.2.1, hinting at a correlation between [C/Fe] in a galaxy and its halo mass. However, C is one of the elements whose surface abundance decreases as low-mass stars evolve, which complicates the interpretation. The correlation might be explained by different distributions of turnoff stars and red giants in kinematics.

While we have shown that most of the elemental abundance ratios [X/Fe] do not show significant correlations with kinematics, the above investigations are limited to comparisons between one-dimensional chemical information and one-dimensional kinematic information. We know that both chemical and kinematics spaces are multidimensional; for example, kinematic substructures cannot be found in a one-dimensional investigation. We thus examined a correlation between multidimensional chemical information and three-dimensional kinematics using the distance correlation defined in Székely et al. (2007).

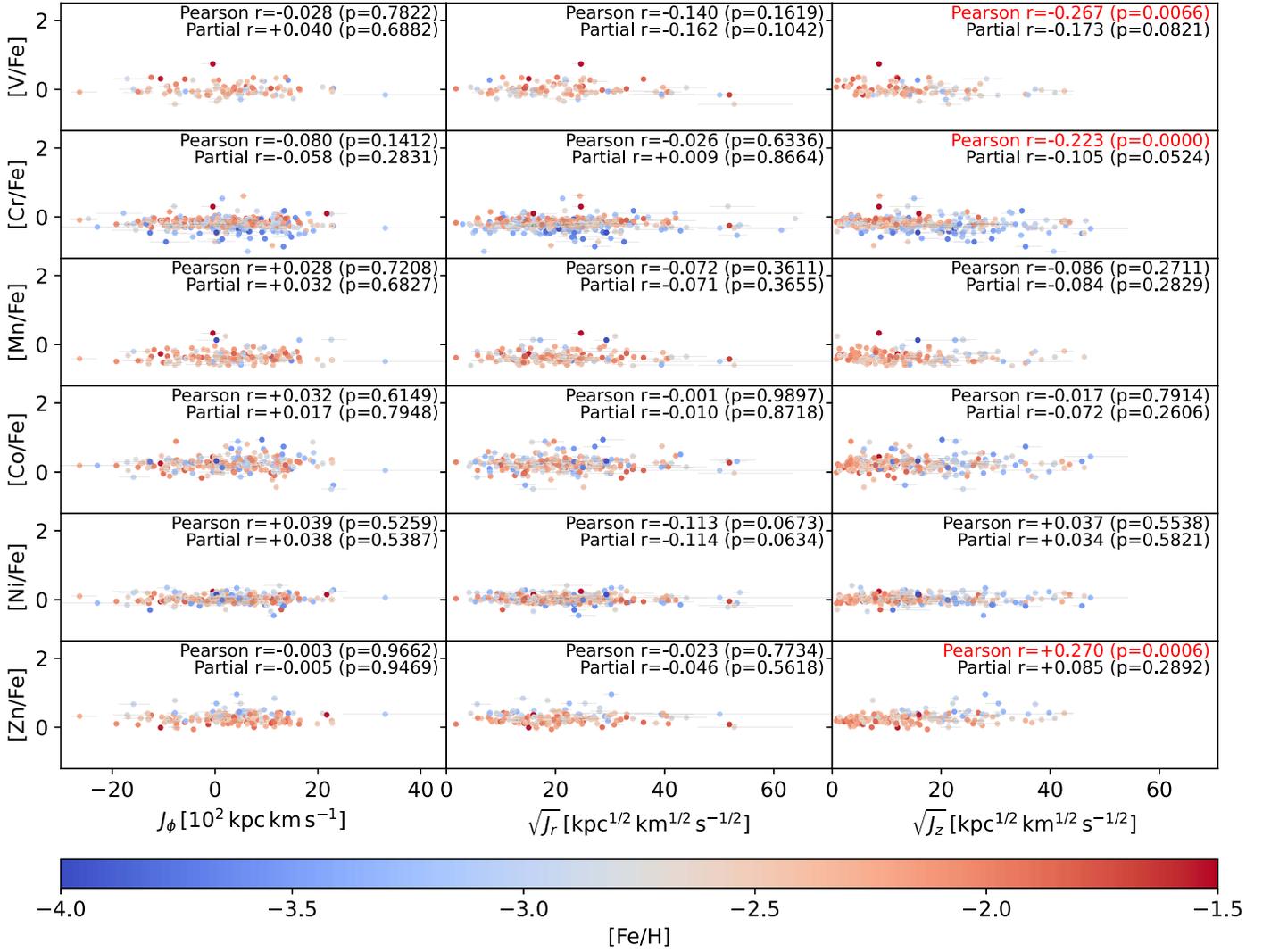

Figure 16. Same as Figure 15, but for iron-group elements.

We first provide a brief summary of distance correlation based on Székely et al. (2007). We considered a distance correlation of $N$ data points between $X$ and $Y$ spaces, which can be multidimensional. Using distance matrices in the $X$ and $Y$ spaces,

$$a_{ij} = |X_i - X_j| \quad (3)$$

$$b_{ij} = |Y_i - Y_j|, \quad (4)$$

where distances are computed in the Euclidean geometry, Székely et al. (2007) defined the following two matrices:

$$A_{ij} = a_{ij} - \frac{1}{n}\sum_k a_{ik} - \frac{1}{n}\sum_l a_{lj} + \frac{1}{n^2}\sum_{k,l} a_{k,l} \quad (5)$$

$$B_{ij} = b_{ij} - \frac{1}{n}\sum_k b_{ik} - \frac{1}{n}\sum_l b_{lj} + \frac{1}{n^2}\sum_{k,l} b_{k,l}. \quad (6)$$

The distance correlation $R_d$ is then defined as the square root of

$$R_d^2(X, Y) = \frac{\frac{1}{n^2}\sum_{i,j} A_{ij} B_{ij}}{\sqrt{\frac{1}{n^2}\sum_{i,j} A_{ij}^2} \sqrt{\frac{1}{n^2}\sum_{i,j} B_{ij}^2}}. \quad (7)$$

The distance correlation takes a value in $0 \leqslant R_d \leqslant 1$ and satisfies $R_d = 0$ if and only if $X$ and $Y$ are independent. When $R_d = 1$, there is a relation between $X$ and $Y$ such that $Y = A + bXC$, where $A$ is a vector, $b$ is a scalar, and $C$ is an orthonormal matrix. The null hypothesis that $X$ and $Y$ are independent can be tested by computing the $R_d$ distribution for samples with the order of $Y$ shuffled.[18]

For the chemical space, we considered eight elemental abundance ratios ([X/Fe], where X are Na, Mg, Ca, Sc, Ti, Cr,

---

[18] The distance correlation can approach 1 even if $X$ and $Y$ are independent when their dimensions are high. In such case, we could use the modified distance correlation by Székely & Rizzo (2013), for which the null hypothesis can be tested by using a $t$-distribution if the dimensions of $X$ and $Y$ are large. Since our data do not have such large dimensions, we stuck to the original distance correlation in Székely et al. (2007).

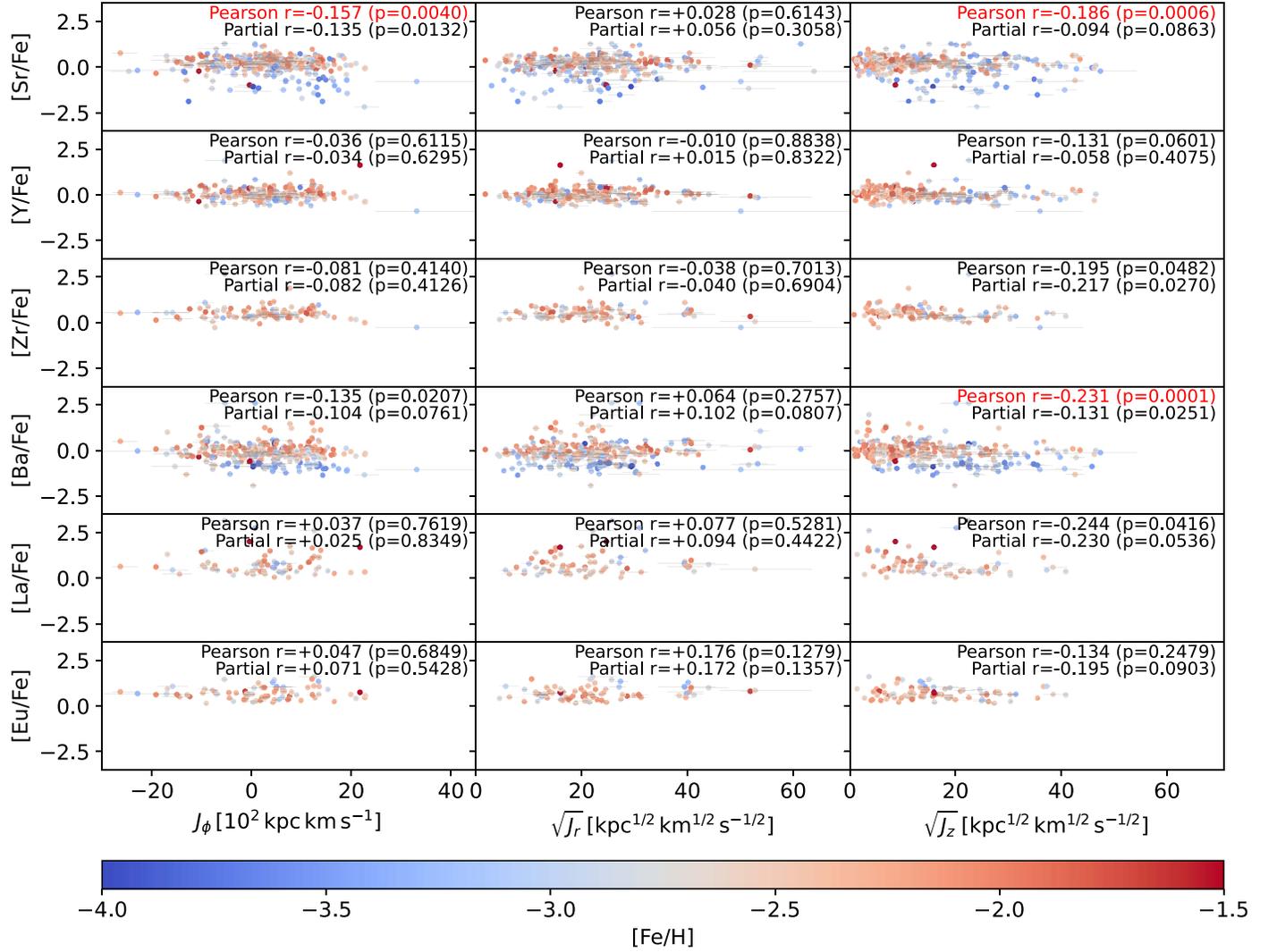

Figure 17. Same as Figure 15, but for neutron-capture elements.

Sr, and Ba) since these elemental abundances are available for a large number of stars. By requiring stars to have all of these ratios, we were left with 273 stars. For the kinematics, we considered the space defined by $J_\phi$, $J_z$, and $J_r$. We further normalized these chemical abundances and kinematics so that the median value became zero and the difference between the 16th and 84th percentiles became 2.

We obtained the distance correlation coefficient of 0.29 with $p < 0.001$, which seems significant. However, we again note that this might simply reflect the metallicity gradient and chemical evolution. Therefore, we ran the same test after removing a linear trend between [X/Fe] and [Fe/H], which yielded the correlation coefficient of 0.23 with $p = 0.07$, suggesting again that the observed significant correlation can be explained by the chemical evolution.

Notably, the absence of a significant correlation here contrasts with the small abundance variations among substructures reported in Section 4.1, such as the lower Zn abundances observed in VMPD. Unlike Section 4.1, which delves into specific kinematic substructures through targeted chemical and kinematic analyses, this section investigates the broader picture across the Milky Way. Here, we focus on identifying global trends in element distributions and stellar motions to understand the overall chemical and dynamical evolution of the Galaxy. While massive accreted galaxies are expected to reside closer to the Milky Way's center due to tidal interactions, our analysis revealed no statistically significant correlation between their chemical composition and kinematics. This suggests that mass may not be the primary driver of early chemical enrichment in these dwarf galaxies.

## 5. Conclusion

In this work, we have conducted a chemodynamical analysis on one of the largest uniform high-resolution VMP star samples, which has been obtained from a joint project between LAMOST and Subaru, i.e., the HR sample. To ensure a robust clustering, the identification of dynamical groups has first been made on the master catalog of the HR sample, which consists of 5778 VMP stars mostly based on the LAMOST low-resolution VMP catalog and is referred to as the LR sample. Dynamical parameters have been estimated using Agama, and the FoF algorithm has been adopted for the clustering. The analysis has resulted in 131 DTGs, including 227 stars from the HR sample, 89 of which have been later associated with several known Galactic substructures, including GSE, Helmi streams, Thamnos, Sequoia, Wukong, and Pontus, as well as a very metal-poor disk-like substructure called VMPD. It thus enabled us to systematically explore the chemical properties of these substructures and dynamical groups down to such a low-metallicity region with [Fe/H] < −2.0. The major results include:

1. About one-fourth of the HR sample belongs to GSE. In general, the distribution of elemental abundances of GSE is similar to that of the whole HR sample and the nonclustered VMP stars, accompanied by large scatters, indicating a massive progenitor system with a complicated enrichment history of GSE.
2. The VMPD shares similar distributions in kinematics and metallicities as the low-eccentricity, prograde, very metal-poor disk components identified in previous studies. We identify a systematically lower [Zn/Fe] compared with the whole HR sample, indicating that the VMPD may have originated from small stellar systems. Such a feature supports the scenario that stars associated with the VMPD have probably originated from low-mass building blocks of the proto-Galaxy.
3. All of the stars in high-energy Sequoia ($E > -1.35 \times 10^5$ km$^2$ s$^{-2}$) are r-I stars with [Eu/Fe] ∼ 0.7, revealing the r-process-enhanced feature of Sequoia at [Fe/H] < −2.
4. Helmi streams show deficiencies in carbon and light neutron-capture elements. This can be explained by the lack of rotating massive stars in the progenitor of Helmi streams. The chemical distribution of other substructures and DTGs is rather similar to the whole HR sample.
5. The fraction of CEMP-no stars seems to vary among different substructures in the HR sample. We find that, in general, the Group II and III CEMP stars in the A(C)−[Fe/H] diagram are more likely to be associated with accreted small dwarf galaxies, which is consistent with previous studies. For GSE and Thamnos, the fractions of CEMP-no stars in both turnoff stars and giants are quite similar to that in nonclustered stars, while for VMPD and the Helmi streams, there are no CEMP-no stars detected in the HR sample. Although such differences can be related to the properties of their progenitor dwarf galaxies, larger samples of VMP stars would be necessary to unravel their origins.
6. There is no clear evidence of whether neutron-capture-element peculiar stars are more likely to be accreted. However, we do find a very interesting subgroup in GSE that exhibits a significant excess of r-process material. Stars in this subgroup provide a great opportunity to explore the complete r-process pattern and, thus, the r-process nucleosynthesis in GSE.
7. It is expected that more massive galaxies sink deeper in the Milky Way potential, which would create a gradient in the progenitor galaxy mass as a function of kinematics. We studied if we can observe the signature of this gradient in abundance ratios by studying large-scale correlations between kinematics and chemical abundances among VMP stars. For most elements, there is no significant correlation, indicating that accreted galaxies of the Milky Way experienced similar chemical evolutions at early times. The only exception is C, which shows a significant correlation with vertical action. However, since the C abundance changes during the evolution of low-mass stars, it is needed to confirm the correlation with a better-controlled, larger sample.

Chemodynamics of a large VMP star sample is crucial to understanding the origin of these ancient stars and could provide important observational constraints on the properties of the early Milky Way and ancient accretions of small dwarf galaxies. Though there have been great advances made in relevant fields, there is still a lack of comprehensive chemodynamical studies on the VMP region of the Galactic halo. Therefore, this paper has made such an effort to conduct a comprehensive chemodynamical study on a large high-resolution spectroscopically analyzed VMP star sample, which provides uniform chemical abundances for over 20 species and thus enables an investigation of the internal chemical properties of various substructures.

We have found a number of interesting features for some substructures; however, in general, we find little significant difference in the abundance trends for different elements between VMP substructures and nonclustered VMP halo stars. This may partly be related to the fact that at such early times with very low metallicities, the chemical enrichment of most galaxies is dominated by nucleosynthesis through CCSNe.

Even though our HR sample is one of the largest VMP star catalogs with homogeneously derived abundances for over 350 objects and we utilized a much larger LR sample with about 5800 VMP stars for the clustering, the number of VMP stars in our sample of the smallest galaxies such as UFDs is insufficient for robust clustering analysis or to observe chemical signatures of individual accreted small galaxies. Future spectroscopic surveys, such as 4MOST, WEAVE, and PFS, will significantly increase the number of VMP stars with abundances of various elements, which shall allow us to ultimately upscale this study.

## Acknowledgments

We thank the anonymous referee for the very efficient and helpful report to improve the draft. This work was supported by the National Natural Science Foundation of China grant Nos. 11988101, 12222305, and 11973049, the National Key R&D Program of China No.2019YFA0405500, the JSPS–CAS Joint Research Program, the CAS Project for Young Scientists in Basic Research (No. YSBR-092), and the science research grants from the China Manned Space Project. This work was also supported by JSPS KAKENHI grant Nos. 20H05855, 21H04499, 22K03688, and 23HP8014. H.L. and Q.X. acknowledge support from the Strategic Priority Research Program of Chinese Academy of Sciences grant No. XDB34020205 and the Youth Innovation Promotion Association of the CAS (id. Y202017 and 2020058). Q.X. acknowledges support from the Beijing Municipal Natural Science Foundation grant No. 1242031. T.M. is supported by a Gliese Fellowship at the Zentrum für Astronomie, University of Heidelberg, Germany.

This research is based on data collected at Subaru Telescope, which is operated by the National Astronomical Observatory of Japan. We are honored and grateful for the opportunity of observing the Universe from Maunakea, which has cultural, historical, and natural significance in Hawaii. Guoshoujing Telescope (the Large Sky Area Multi-Object Fiber Spectroscopic Telescope, LAMOST) is a National Major Scientific Project built by the Chinese Academy of Sciences. Funding for the project has been provided by the National Development and Reform Commission. It is operated and managed by the National Astronomical Observatories, Chinese Academy of Sciences. This work has made use of data from the European Space Agency (ESA) mission Gaia (https://www.cosmos.esa.int/gaia), processed by the Gaia Data Processing and Analysis

Consortium (DPAC, https://www.cosmos.esa.int/web/gaia/dpac/consortium). Funding for the DPAC has been provided by national institutions, in particular the institutions participating in the Gaia Multilateral Agreement.

*Facilities:* LAMOST, Subaru

*Software:* Agama (Vasiliev 2019), TOPCAT (Taylor 2005).

## Appendix A
## LR and HR Sample Distance Distribution

The distance distributions of stars from the reduced L18 catalog and the HR sample are shown in Figure A1.

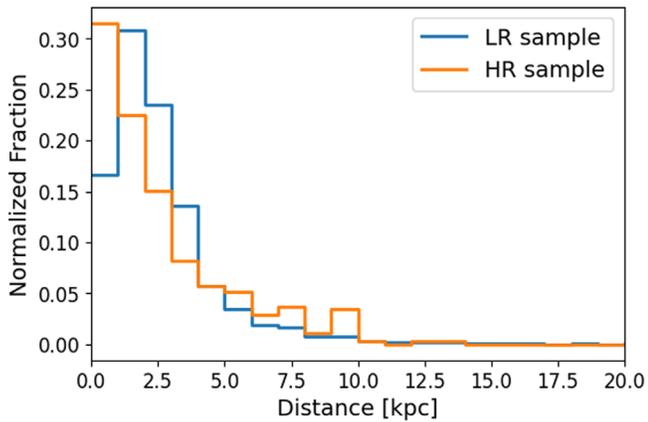

**Figure A1.** The normalized distance distribution fraction of stars from the reduced L18 catalog and the HR sample. The bin size of these two samples is 1 kpc. Stars in the HR sample show a larger fraction of stars located farther than 5 kpc; this may be caused by the selection bias from the selection strategy (see Section 2.2 of Paper I).

## Appendix B
## Determination of the Linking Length

We use the group to which J2216+0246 belongs as an example (hereafter example group) to clarify the choice of linking length in the FoF algorithm. Figure B1 shows how the size of the example group changes as the linking length increases. In most cases, the group size initially grows slowly and then jumps up at a certain linking length (e.g., 0.094 and 0.100 for the example group). Since the rapid increase could correspond to a merger of two or more groups, we checked if the group shows a multimodal distribution after the jump. We adopted the linking length before the jump, which leads to the formation of a multimodal DTG. As shown in Figure B2, the example group clusters well at a linking length of 0.098 but shows three subcomponents at 0.100, indicating that the merger at 0.100 connects multiple DTGs. Therefore, we took 0.098 as the linking length of this DTG.

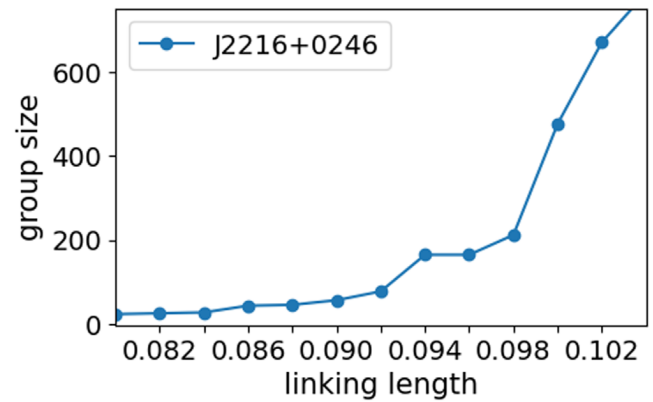

**Figure B1.** The size of the example group at different linking lengths.

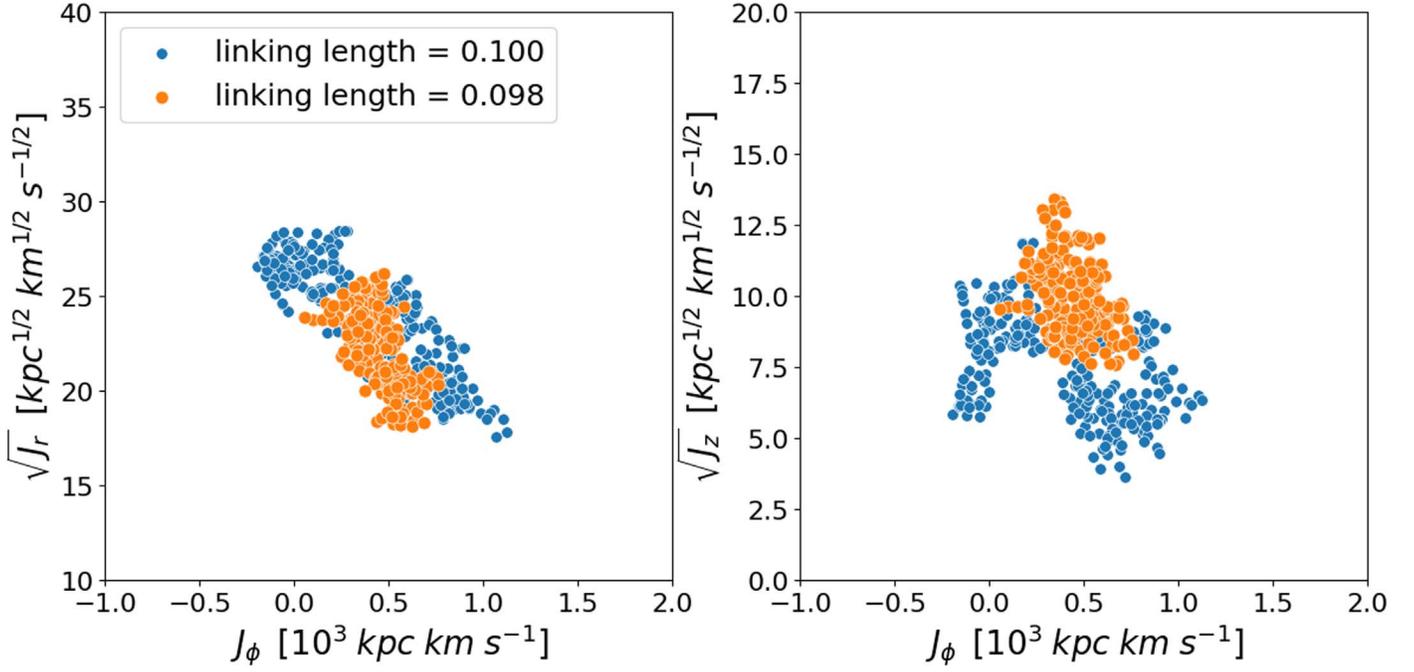

**Figure B2.** Distribution in clustering phase space of the example group at linking length 0.098 (orange) and 0.100 (blue).

## Appendix C
## Substructure Criteria

The distribution of eccentricity and $|z|_{max}$ for prograde LR sample stars is shown in Figure C1. We selected the VMPD members based on the overdensity within the red circle in Figure C1. Table C1 shows the criteria we adopted to associate DTGs with different substructures.

**Table C1**
Criteria Adopted for Each Substructure

| Substructures | Criteria | References |
|---|---|---|
| GSE | $ec > 0.7$; $|v_\phi| < 100$ (km s$^{-1}$) | (1, 2) |
| Thamnos | $\eta < -0.4$; $-1.8 < E < -1.6$ ($\times 10^5$ km$^2$ s$^{-2}$) | (1, 3) |
| Sequoia | $\eta < -0.15$; $J_\phi < -0.7$ ($\times 10^3$ kpc km s$^{-1}$); $E > -1.5$ ($\times 10^5$ km$^2$ s$^{-2}$) | (1, 4) |
| Helmi streams | $0.75 < J_\phi < 1.7$ ($\times 10^3$ kpc km s$^{-1}$); $1.6 < L_\perp < 3.2$ ($\times 10^3$ kpc km s$^{-1}$) | (1, 5) |
| Wukong (LMS-1) | $0 < J_\phi < 1.0$ ($\times 10^3$ kpc km s$^{-1}$); $E < -1.15$ ($\times 10^5$ km$^2$ s$^{-2}$); $(J_z - J_r)/J_{tot} > 0.3$; $90° < \arccos(L_z/L) < 120°$ | (6, 7) |
| Pontus | $-1.72 < E < -1.56$ ($\times 10^5$ km$^2$ s$^{-2}$); $-470 < J_\phi < 5$ (kpc km s$^{-1}$); $245 < J_r < 725$ (kpc km s$^{-1}$); $115 < J_z < 545$ (kpc km s$^{-1}$); $390 < L_\perp < 865$ (kpc km s$^{-1}$); $0.5 < ec < 0.8$; $1 < r_{peri} < 3$ (kpc); $8 < r_{apo} < 13$ (kpc) | (8, 9) |
| VMPD | $0.25 < ec < 0.6$; $|z|_{max} < 3$ (kpc); $(v_y - 233.1)^2 + v_x^2 + v_z^2 < 180$ (km s$^{-1}$); $v_\phi > -200$ (km s$^{-1}$) | (10) |

**References.** (1) Naidu et al. (2020); (2) Belokurov et al. (2018); (3) Koppelman et al. (2019a); (4) Myeong et al. (2019); (5) Koppelman et al. (2019b); (6) Limberg et al. (2024); (7) Yuan et al. (2020a); (8) Horta et al. (2023); (9) Malhan (2022); (10) see Section 4.1.2 and Figure C1.

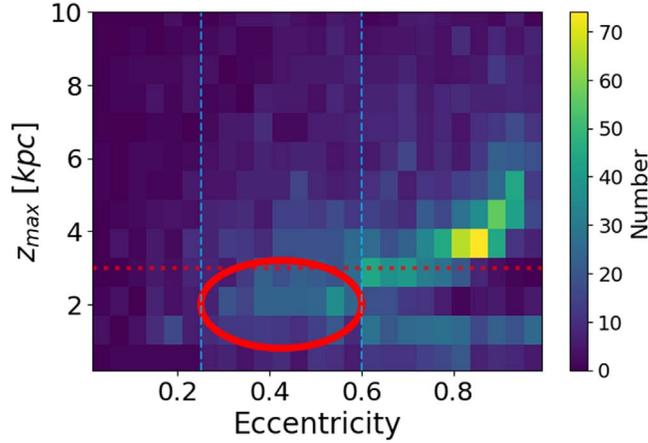

**Figure C1.** The distribution of eccentricity and $|z|_{\max}$ for prograde LR sample stars. We can find an overdensity within the red circle with $0.25 < ec < 0.60$ and $|z|_{\max} < 3$ (kpc) on this panel. The red dotted line is $|z|_{\max} = 3$ (kpc) and blue dashed lines are $ec = 0.25$ and $ec = 0.60$.

## Appendix D
## Comparison with Literature Results

To compare our results with literature and high-resolution sky survey catalogs, we collected the published chemical abundances for GSE, Thamnos, Sequoia, Helmi streams, and Wukong and summarize the results in Table D1. In addition, we used the stars from the high-$\alpha$ sequence in Nissen & Schuster (2010) and stars not classified as GSE in Reggiani et al. (2017) as background stars. For each work, we converted the chemical abundances to that under the solar abundances from Asplund et al. (2009).

We also made use of the GALAH DR3 catalog (Buder et al. 2021), which could provide us with 16 common elements to compare with our results. Following the same procedure as described in Section 2.2, we reduced the GALAH DR3 catalog and computed the dynamic parameters for stars within 4 kpc. We then applied the following cut for the GALAH DR3 nearby sample to remove stars with unreliable measurements and exclude the observations of GCs and the bulge (Buder et al. 2021, 2022):

1. flag_sp=0 and flag_fe_h=0,
2. survey_name ≠ "other",
3. 3500 K < $T_{\rm eff}$ < 6250 K.

Subsequently, we used $|v - v_{\rm LSR}| > 180$ km s$^{-1}$ and [Fe/H] $< -0.8$ to select the halo stars, and applied the criteria in Table C1 to select Thamnos, Sequoia, Helmi streams, and Wukong. For GSE, we added the criterion $\sqrt{J_r} > 30$ kpc km s$^{-1}$ to avoid the contamination of the heated disk (see, e.g., Feuillet et al. 2020; Matsuno et al. 2021; Carrillo et al. 2024). We also identified the prograde stars with $\sqrt{J_r} < 15$ kpc km s$^{-1}$ as in situ components (see Matsuno et al. 2021) for comparison with the accreted substructures. We note that this

**Table D1**
Chemical Results from Literature for Each Substructure

| Substructures | Literature | Common Elements | $N_{\rm Stars}$ |
|---|---|---|---|
| GSE | NS10&11[a,b] | Na I, Mg I, Si I, Ca I, Ti II, Cr I, Mn I, Ni I, Zn I, Y II, Ba II | 16 |
|  | Reggiani et al. (2017)[b] | Na I, Mg I, Si I, Ca I, Ti II, Cr I, Mn I, Ni I, Zn I, Y II, Ba II | 6 |
|  | Monty et al. (2020) | Na I, Mg I, Si I, Ca I, Ti II, Cr I, Ni I, Y II, Ba II | 5 |
|  | Aguado et al. (2021a) | C, Sr II, Y II, Ba II, Eu II | 4 |
|  | Carrillo et al. (2022) | Na I, Mg I, Si I, Ca I, Sc II, V I, Cr I, Mn I, Co I, Ni I, Zn I, Y II, Zr II, Ba II, La II, Eu II | 62 |
| Thamnos | NS10&11[a,b] | Na I, Mg I, Si I, Ca I, Ti II, Cr I, Mn I, Ni I, Zn I, Y II, Ba II | 7 |
|  | Monty et al. (2020)[c] | Na I, Mg I, Si I, Ca I, Ti II, Cr I, Ni I, Y II, Ba II | 3 |
| Sequoia | Monty et al. (2020)[c] | Na I, Mg I, Si I, Ca I, Ti II, Cr I, Ni I, Y II, Ba II | 3 |
|  | Aguado et al. (2021a) | C, Sr II, Y II, Ba II, Eu II | 5 |
|  | Matsuno et al. (2022b) | Na I, Mg I, Si I, Ca I, Ti II, Cr I, Mn I, Ni I, Zn I, Y II, Ba II | 12 |
| Helmi streams | Roederer et al. (2010) | C, Mg I, Si I, Ca I, Sc II, Ti II, V I, Cr I, Mn I, Co I, Ni I, Zn I, Sr II, Y II, Zr II, Ba II, La II, Eu II | 12 |
|  | Aguado et al. (2021b) | C, Na I, Mg I, Si I, Ca I, Sc II, Ti II, V I, Cr I, Mn I, Co I, Ni I, Sr II, Ba II | 7 |
|  | Gull et al. (2021) | C, Na I, Mg I, Si I, Ca I, Sc II, Ti II, Cr I, Mn I, Co I, Ni I, Zn I, Sr II, Y II, Zr II, Ba II, La II, Eu II | 12 |
|  | Matsuno et al. (2022a) | Na I, Mg I, Si I, Ca I, Ti II, Cr I, Mn I, Ni I, Zn I, Y II, Ba II | 11 |
| Wukong | Limberg et al. (2024) | C, Na I, Mg I, Si I, Ca I, Sc II, Ti II, V I, Cr I, Mn I, Co I, Ni I, Zn I, Sr II, Y II, Zr II, Ba II, La II, Eu II | 14 |

**Notes.**
[a] NS10&11 refers to the abundances from Nissen & Schuster (2010) and Nissen & Schuster (2011).
[b] We used the homogenized abundances from Matsuno et al. (2022b) for these two samples. The GSE sample is kinematically selected by Matsuno et al. (2022b), and the Thamnos sample comprises the low-[Fe/Mg] stars in Nissen et al. (2024), which are suggested to belong to Thamnos.
[c] The "SeqG1" and "SeqG2" in Monty et al. (2020) are considered as Thamnos and Sequoia in this work, respectively.

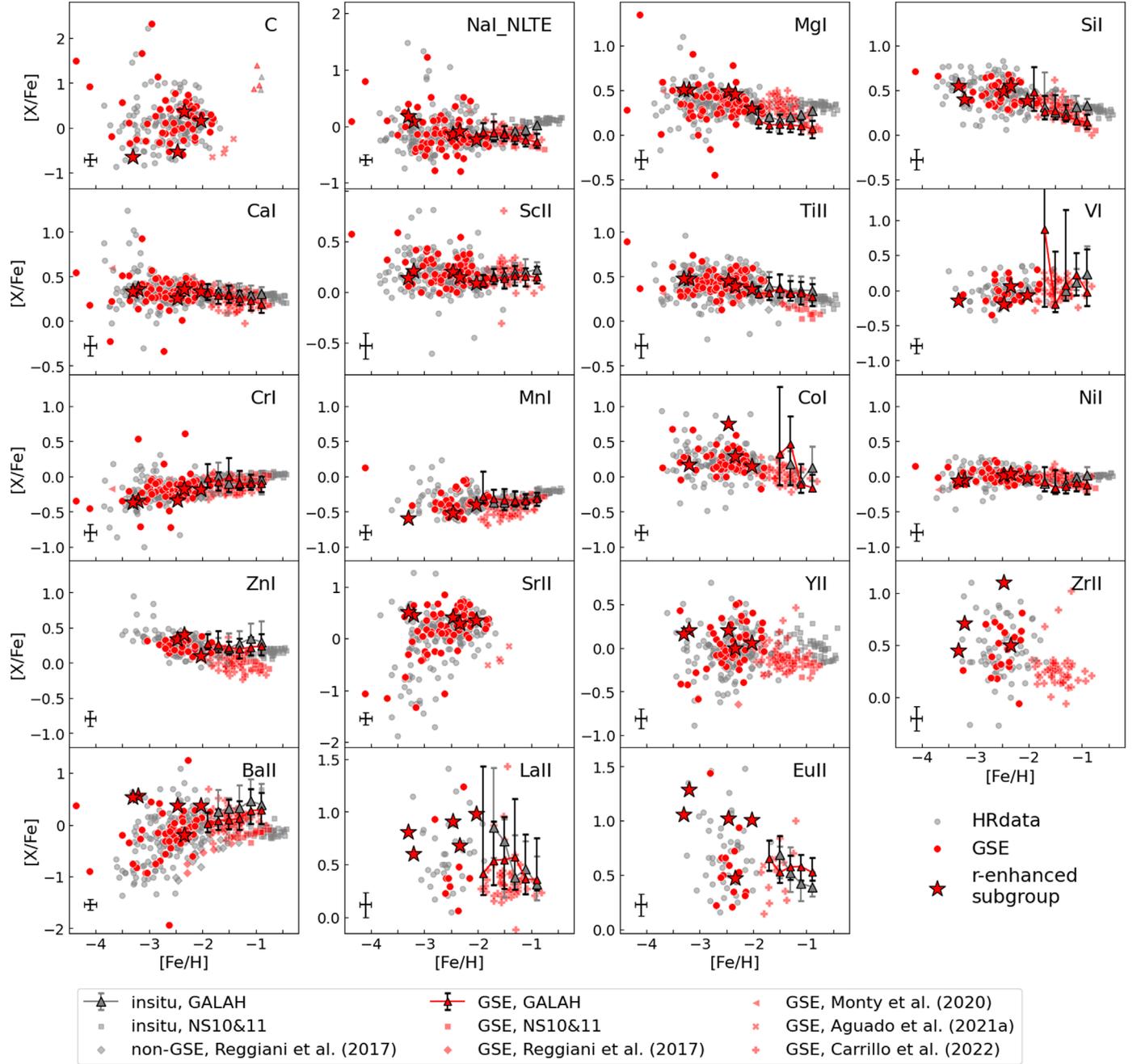

**Figure D1.** The combined chemical distributions of GSE. The colors show different substructure memberships, and the shapes show different sources of abundances. The gray symbols are background stars, combining all stars with reliable abundances in this work (circles), high-α-sequence stars in Nissen & Schuster (2010) and Nissen & Schuster (2011; squares), stars not classified as GSE in Reggiani et al. (2017; diamonds), and in situ components selected from GALAH DR3 (triangles with error bars). The red symbols are GSE members. Results from this paper are presented as circles, and the other symbols represent results from Nissen & Schuster (2010) and Nissen & Schuster (2011; squares), Reggiani et al. (2017; diamonds), Monty et al. (2020; left triangles), Aguado et al. (2021a; filled Xs), and Carrillo et al. (2022; filled pluses). The GALAH data (triangles) are binned in metallicity, and the median values are plotted. The upper and lower error bars are the 84th and 16th percentiles in each bin. For elements with fewer than three stars in each of these bins, we plot them directly in the corresponding panels. The meaning of the rest of the symbols is the same as in Figure 7.

simple selection method would contaminate the substructure samples, so we primarily focused on trends in the distribution of the GALAH sample rather than on the abundance properties of individual stars.

Similar to Section 4.1, we only kept stars with uncertainties less than 0.2 dex in [Fe/H] and [X/Fe] during discussions for corresponding element X. The combined chemical distribution of GSE, retrograde substructures (Thamnos and Sequoia), and polar substructures (Helmi streams and Wukong) are shown in Figures D1, D2, and D3, respectively. The colors show different substructure membership, and the shapes show different sources of abundance.

In general, the chemical distribution of these substructures is similar to that in the discussion in Section 4.1 and previous studies. For example, we can confirm the moderate r-process enhancement of Sequoia members in Figure D2, and the light

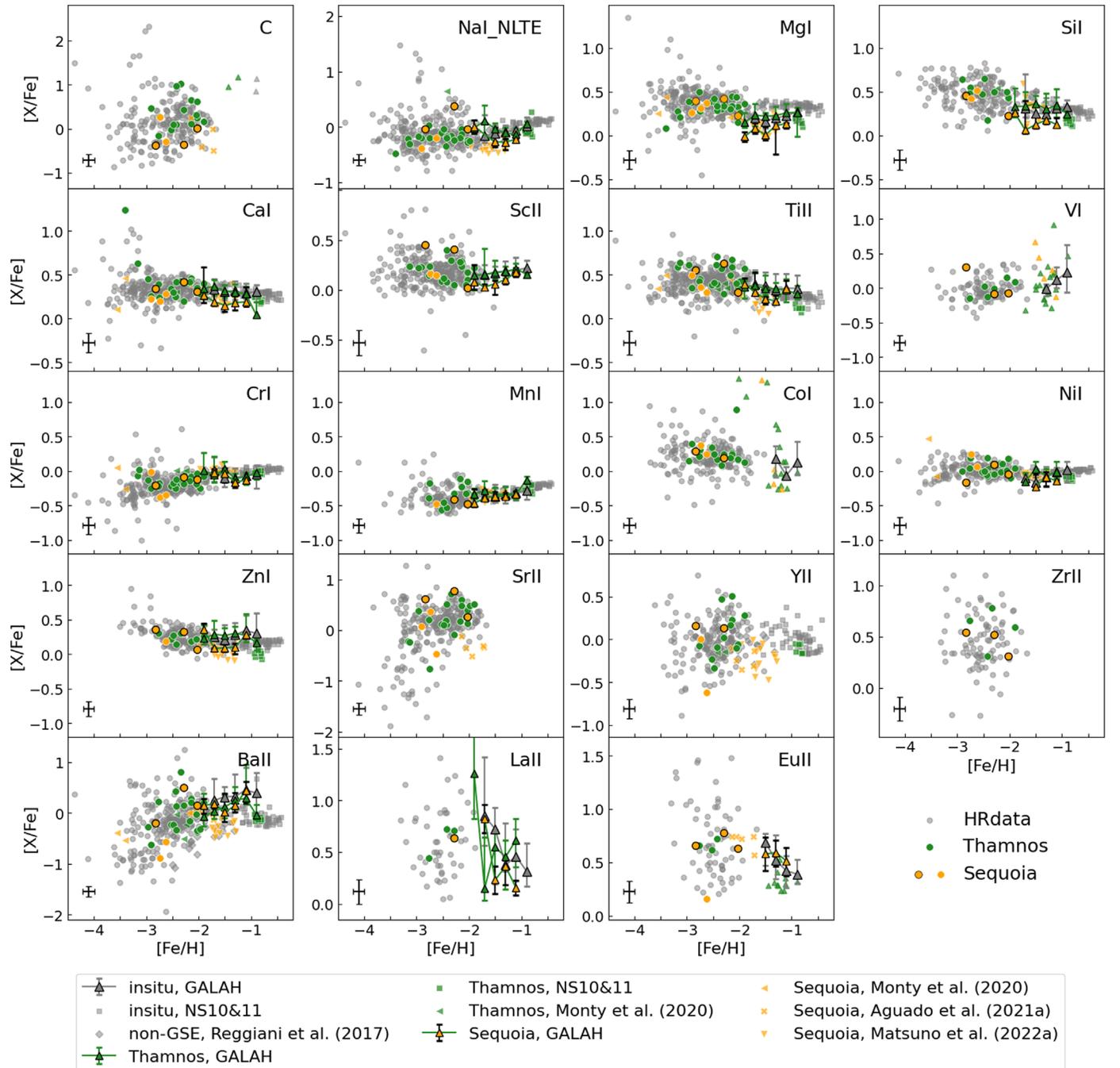

**Figure D2.** The combined chemical distributions of Thamnos and Sequoia. The green and yellow symbols are members of Thamnos and Sequoia, respectively. The meaning of the gray symbols and triangle symbols (GALAH data) is the same as in Figure D1. Results from this paper are presented as circles, and the Sequoia members are divided into two parts, presented by circles with black borders ($E > -1.35 \times 10^5$ km$^2$ s$^{-2}$) and with white borders ($E < -1.35 \times 10^5$ km$^2$ s$^{-2}$). The other symbols represent results from Nissen & Schuster (2010) and Nissen & Schuster (2011; squares), Monty et al. (2020; left triangles), Aguado et al. (2021a; filled Xs), and Matsuno et al. (2022b; inverted triangles). The meaning of the rest of the symbols is the same as in Figure 9.

neutron-capture elements (e.g., Sr, Zr) deficiency of Helmi stream members in Figure D3.

However, we note that directly combining the chemical abundances from the literature without homogenization would introduce systematic uncertainties. The systematic uncertainties can be significant and affect our interpretation of the chemical properties (e.g., see Section 4.2 in Matsuno et al. 2022a). The differences between the GALAH DR3 in situ sample and Nissen & Schuster (2010) in situ members for several elements (e.g., Mg, Ba, etc.) show the effect of systematic uncertainties in our combined sample. Therefore, chemodynamical analyses based on large, uniform, high-resolution samples analyzed using a homogeneous method could provide more reliable chemical properties of the substructures.

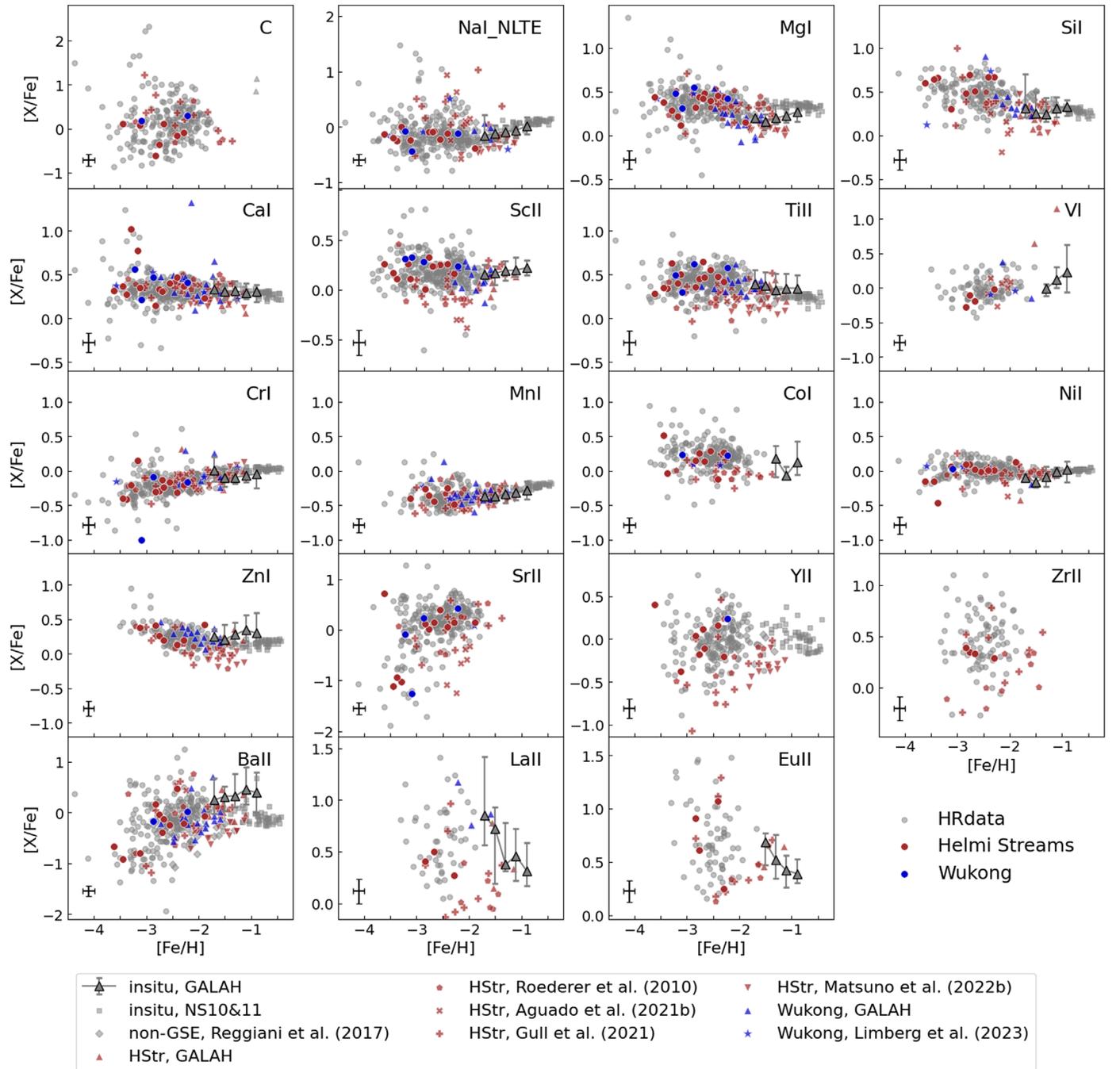

**Figure D3.** The combined chemical distributions of Helmi streams and Wukong. The brown and blue symbols are members from Helmi streams and Wukong, respectively. The meaning of the gray symbols and triangle symbols (GALAH data) is the same as in Figure D1. Results from this paper are presented as circles, and the other symbols represent results from Roederer et al. (2010; pentagons), Aguado et al. (2021b; filled Xs), Gull et al. (2021; filled pluses), Matsuno et al. (2022a; inverted triangles), and Limberg et al. (2024; stars). The meaning of the rest of the symbols is the same as in Figure 10.

**ORCID iDs**

Ruizhi Zhang ⓘ https://orcid.org/0009-0008-1319-1084
Tadafumi Matsuno ⓘ https://orcid.org/0000-0002-8077-4617
Haining Li ⓘ https://orcid.org/0000-0002-0389-9264
Wako Aoki ⓘ https://orcid.org/0000-0002-8975-6829
Xiang-Xiang Xue ⓘ https://orcid.org/0000-0002-0642-5689
Takuma Suda ⓘ https://orcid.org/0000-0002-4318-8715
Gang Zhao ⓘ https://orcid.org/0000-0002-8980-945X
Yuqin Chen ⓘ https://orcid.org/0000-0002-8442-901X
Miho N. Ishigaki ⓘ https://orcid.org/0000-0003-4656-0241
Jianrong Shi ⓘ https://orcid.org/0000-0002-0349-7839
Qianfan Xing ⓘ https://orcid.org/0000-0003-0663-3100
Jingkun Zhao ⓘ https://orcid.org/0000-0003-2868-8276